\documentclass[aps,showpacs,nofootinbib,amssymb]{revtex4}
\usepackage{graphicx,epsfig}
\usepackage{bm}
\usepackage{amsfonts,amssymb}
\usepackage{xcolor}
\usepackage[pdftex]{hyperref} 

\newcommand{\bG}{{\beta_G}}

\renewcommand{\bG}{B_\rmi{G}}    
\newcommand{\fr}[2]{{\frac{#1}{#2}\,}}
\newcommand{\rmi}[1]{{\mbox{\scriptsize #1}}}

\newcommand{\nn}{\nonumber}
\newcommand{\be}{\begin{equation}}
\newcommand{\ee}{\end{equation}}
\newcommand{\bea}{\begin{eqnarray}}
\newcommand{\eea}{\end{eqnarray}}
\def\al{\alpha}

\def\siml{{\ \lower-1.2pt\vbox{\hbox{\rlap{$<$}\lower6pt\vbox{\hbox{$\sim$}}}}\ }} 
\def\lQ{\Lambda_{\rm QCD}}

\newcommand{\MS}{\overline{\rm MS}}

\newcommand{\ord}{{\cal O}}

\def\msb{{\overline{\rm MS}}}

\begin{document}
\preprint{UAB-FT-XXX}

\title{\boldmath 
The QCD static potential in $D<4$ dimensions at weak coupling
\unboldmath}
\author{Antonio Pineda\footnote{Electronic address: pineda@ifae.es} and Maximilian Stahlhofen\footnote{Electronic address: stahlhofen@ifae.es}}
\affiliation{Grup de F\'\i sica Te\`orica and IFAE, Universitat
Aut\`onoma de Barcelona, E-08193 Bellaterra, Barcelona, Spain
}
\date{\today}

\begin{abstract}
\noindent
We study the static potential of a color singlet quark-antiquark pair 
with (fixed) distance $r$ in $D=3$ and $D=2$ space-time dimensions at weak coupling 
($\al\, r \ll 1$ and $g \,r \ll 1$, respectively). Using the effective theory 
pNRQCD we determine the ultrasoft contributions, which cannot be computed in 
conventional perturbative QCD. We show in detail how the ultrasoft 
renormalization in pNRQCD is carried out. In three dimensions the 
precision of our results reaches ${\cal O}(\al^3r^2)$, i.e. NNLO in 
the multipole expansion, and NNLL in a $\al/\Delta V$ expansion, where 
$\Delta V \sim \al \ln(\al r)$. We even present results up to partly $\rm N^4$LL order 
and compare them to existing lattice data. Finally we discuss the relevance of the 
perturbative calculation in two dimensions, where the exact result is known.

\end{abstract}
\maketitle

\section{Introduction}

The singlet potential between two color sources in the static limit, i.e. with infinite masses, 
is one of the most accurately studied objects in QCD. It plays an important role in understanding the 
dynamics of QCD. On the one hand it is an essential ingredient in a 
Schr\"odinger-like description of heavy quarkonia. On the other hand, a linear growing potential at long distances (in the large 
$N_c$ limit) indicates confinement of the strong interactions. This has motivated 
a lot of effort to steadily increase the accuracy of the perturbative prediction for 
that object~\cite{FSP,Schroder:1999sg,short,KP1,RG,Hoang:2002yy,Brambilla:2006wp,Brambilla:2009bi,Anzai:2009tm,Smirnov:2009fh}, 
which should be reliable at short distances. A crucial point is obviously to distinguish which region can be described within perturbation 
theory and to understand the crossover to the non-perturbative regime~\cite{Recksiegel:2001xq,Pineda:2002se,Lee:2002sn},~\cite{Brambilla:2009bi}. 

Besides the singlet also the color octet static potential has attracted some interest in the last decade. 
It has been computed with two-loop precision at weak coupling in Ref.~\cite{Kniehl:2004rk}. Together with 
the so-called gluelumps, it describes the short-distance behavior of 
the hybrid potentials~\cite{Brambilla:1999xf,Bali:2003jq}, which are potentially important for the theoretical prediction 
of physical hybrids made of heavy quarks. At long distances the hybrid potentials could also help 
to shed light on the question whether the dynamics responsible for confinement is of string type and if so, 
to determine its structure exactly. 

In this paper, we will focus on the singlet static potential, $V_s(r)$, 
and on the singlet static energy, $E_s(r)$, at short distances. $V_s(r)$ is the leading order
potential introduced in a Schr\"odinger-like description of heavy quarkonia.
$E_s(r)$ is the energy of a static color singlet quark-antiquark pair with (fixed) distance $r$, 
being the object actually computed on  the lattice. 
Its physics at short distances is governed by at least two physical scales.
One is the soft scale $\sim 1/r$, and the other one is the ultrasoft scale $\sim V_s$. 
The most convenient framework to perform the corresponding calculations is the effective field theory (EFT)
``potential NRQCD'' (pNRQCD) \cite{Pineda:1997bj} (for a review see \cite{Brambilla:2004jw}). 
pNRQCD also has been instrumental in many of the latest perturbative results 
quoted above, especially in the determination of the ultrasoft contributions.
In its static limit the theory exploits the scale separation $1/r \gg V_s$. $V_s$ 
accounts only for the soft scale contributions to the static energy and can be computed in 
perturbative QCD (pQCD). $E_s$ also includes effects at the ultrasoft scale, which in the 
static limit is generated by a resummation of certain classes of loop diagrams to all 
orders and serves as an infrared (IR) regulator~\cite{Appelquist:1977es}. These contributions 
can be calculated in pNRQCD from a finite number of diagrams.

In order to gain a deeper understanding of the static potential one can consider how it 
is qualitatively affected by changing the number of dimensions from four (4D) to three (3D) or two (2D). 
The three-dimensional result is also important on its own. Four dimensional thermal QCD 
undergoes effectively a dimensional reduction for large temperatures. 
Therefore, determining the renormalization group (RG) structure for the static potential in 
three dimensions may open the way to a resummation of logarithms at finite temperature. 
Three dimensional space-time is moreover a good testing ground for renormalon issues, 
since the linear power divergences associated with renormalons in four dimensions become logarithmic divergences 
in three dimensions and can be traced back using dimensional regularization. 
The computations in three and two dimensions finally represent consistency 
checks of the theoretical approach used to describe the 4D potential. 

Partially driven by these motivations the computation of the static potential has been 
carried out up to two loops for arbitrary dimension $D$ in Ref.~\cite{Schroder:1999sg}. Whereas in four 
dimensions the result is finite, this is not the case in three dimensions. 
Beyond one-loop it suffers from IR singularities, whose origin was obscure at the time they were encountered.
In particular, it was not clear whether the space-like end-strings of the rectangular Wilson loop, which 
defines the static potential non-perturbatively, could contribute, raising doubts about the 
independence of the result on the precise form of the strings.

In 4D such IR singularities first appear at three-loop level and were shown to cancel with the ultrasoft ultraviolet (UV)
singularities in Ref.~\cite{short}. Up to now an analogous ultrasoft computation has been missing in three dimensions. Therefore, 
the precision of the 3D static potential has been limited to ${\cal O}(\al^2 r)$ from the finite tree-level and one-loop results.
With this work we will show for the color singlet state that the UV divergences 
of the ultrasoft pNRQCD calculation in fact cancel the soft pQCD IR singularities exactly. 
This allows us, for the first time, to go beyond the one-loop level and to obtain 
the logarithmic corrections at two loops as well as the full analytical structure of 
the $\ord(r^2)$ correction. We will point out the peculiarities of the 3D computation in detail. 

The outline of the paper is as follows.
In Sec.~\ref{pNRQCD} we introduce the theoretical setup for our calculations, 
we define the power counting of the EFT in $D<4$ dimensions and review existing 
results for the relevant ultrasoft loop diagrams. Sec.~\ref{sec3D} contains the 
complete RG improved ultrasoft calculation up to NNLL and partly up to $\rm N^4$LL 
order in three dimensions as well as a comparison of the results to lattice data. 
In Sec.~\ref{secD2} we discuss the situation in two dimensions and in Sec.~\ref{conclusion} we present our conclusions.

\section{pNRQCD}
\label{pNRQCD}

Up to the next-to-leading order (NLO) in the multipole expansion (and irrespectively of the space-time dimension) 
the effective Lagrangian density of pNRQCD in the static limit
 takes the form~\cite{Pineda:1997bj},~\cite{Brambilla:1999xf}:
\begin{eqnarray}
& & {\cal L}_{\rm us} =
{\rm Tr} \Biggl\{ {\rm S}^\dagger \left( i\partial_0  - V_s(r)   \right) {\rm S} 
 + {\rm O}^\dagger \left( iD_0 - V_o(r)  \right) {\rm O} \Biggr\} 
\nonumber\\
& &\qquad
+ g V_A ( r) {\rm Tr} \left\{  {\rm O}^\dagger {\bf r} \cdot {\bf E} \,{\rm S}
+ {\rm S}^\dagger {\bf r} \cdot {\bf E} \,{\rm O} \right\} 
  + g {V_B (r) \over 2} {\rm Tr} \left\{  {\rm O}^\dagger \left\{{\bf r} \cdot {\bf E} , {\rm O}\right\}\right\} +  \ord(r^2)\,.
\label{pnrqcd0}
\end{eqnarray}
We define color singlet and octet fields for the quark-antiquark system by $S = S({\bf r},{\bf R},t)$ and 
$O^a =  O^a({\bf r},{\bf R},t)$, respectively. ${\bf R} \equiv ({\bf x}_1+{\bf x}_2)/2$ is the center position of the system.
In order for $S$ and $O^a$ to have the proper free-field normalization in color space they are related to the fields in Eq.~(\ref{pnrqcd0}) as follows:
\begin{equation}
{\rm S} \equiv { 1\!\!{\rm l}_c \over \sqrt{N_c}} S\,, \qquad {\rm O} \equiv  { T^a \over \sqrt{T_F}}O^a, 
\label{norm}
\end{equation}
where $T_F=1/2$ for the fundamental representation of SU($N_c$).
All gluon and scalar fields in Eq. (\ref {pnrqcd0}) are evaluated 
in ${\bf R}$ and the time $t$, in particular the chromoelectric field ${\bf E} \equiv {\bf E}({\bf R},t)$ and the ultrasoft covariant derivative
$iD_0 {\rm O} \equiv i \partial_0 {\rm O} - g [A_0({\bf R},t),{\rm O}]$. 
$V_{\{s,o,A,B\}}(r)$ are the Wilson coefficients of the effective Lagrangian. They are fixed 
 at a scale $\nu$ smaller than (or similar to) $1/r$ and larger than the ultrasoft and any other scale in the problem by 
 matching the effective and the underlying theory, which in this case is QCD in the static limit.

\subsection{Power counting}

Because the mass dimension of the coupling is $[g^2]=M^{4-D}$, we have (at least) the following physical scales involved in the problem:
\be
k \sim 1/r \;({\rm soft})\,,\qquad k \sim V \;({\rm ultrasoft})\,, \qquad k \sim g^{\frac{2}{4-D}} \;(\text{non-perturbative})\,.
\ee
In order for perturbation to make sense at the (soft) matching scale we demand 
\be
\label{MP}
gr^{\frac{4-D}{2}} \ll 1
\,,\ee
i.e. weak coupling.
Therefore $g^2r^{4-D}$ plays the role of a dimensionless expansion parameter. 
It is a good expansion parameter at short distances for $D<4$ 
(for $D=4$ Eq.~(\ref{MP}) implies $g \ll 1$). 

We are left with two scales: $V \sim 1/r \times g^2r^{4-D}$ and $g^{\frac{2}{4-D}}$ much smaller than $1/r$ 
as we go to short distances. 
In principle, this guarantees that the multipole expansion makes sense, but 
it says nothing about the relative size of $V$ and $g^{\frac{2}{4-D}}$, i.e. whether one can use 
perturbation theory at the ultrasoft scale $V$. On the other hand note that
\be
V/g^{\frac{2}{4-D}} \sim r^{3-D}g^{2(\frac{3-D}{4-D})} \sim (rg^{\frac{2}{4-D}})^{3-D}
\,.
\ee 
Therefore, $D=3$ is the turning point for the use of perturbation theory at the ultrasoft scale. 

For $4>D>3$ we have a perturbative expansion in $g$:
\be
1/r \gg V \gg g^{\frac{2}{4-D}}
\,.
\ee 
In this case there is a double expansion with the two parameters 
\be
g^2r^{4-D} \ll 1\,, \qquad \frac{g^{\frac{2}{4-D}}}{V} \ll 1\,.
\ee

For $D<3$ the right expansion parameter would be $V/g^{\frac{2}{4-D}}$. 
In principle, this is not a real problem. We just have to expand the exponent in the 
ultrasoft correlators, as we will see later, while the multipole expansion is still valid. 
This is potentially quite interesting. In two dimensions the exact result is 
known and equivalent to the tree-level result. Therefore, strong cancellations are supposed to occur 
among the contributions from the different scales: soft, ultrasoft and $g$. We will further discuss this issue in 
Sec. \ref{secD2}. 

The case $D=3$, the main focus of this paper, deserves a special discussion. 
At short distances we find $V \sim g^2\ln(r\nu)$. This implies that 
\be
g^2/V_s \sim 1/\ln(r\nu) \ll 1\,,
\ee
if $\nu \sim  V$
and we conclude that we can use perturbation theory at the ultrasoft scale $V$. We therefore formally distinguish between 
the scale $V$ and $g^2$. Logarithms from the ultrasoft perturbative computation will have the form $\ln ( V/\nu)$ and are rendered small, if we set $\nu \sim  V$. 
Thus it is legitimate to consider the ultrasoft regime as perturbative, i.e. the pNRQCD loop expansion makes sense (for sufficiently small $r$). 
The inclusion of non-perturbative effects will be discussed in subsection~\ref{npsubsec}.

\subsection{Bare data}

Here we summarize some bare results that will be relevant in the next sections.
In the following we will use the index ``$B$'' to explicitly denote bare quantities. Parameters without this index are understood to be renormalized. 

The general expression for the bare singlet potential in momentum space, ${\tilde V}_{s,B}$, in $D$ dimensions can be written as
\be
{\tilde V}_{s,B}=-C_Fg_B^2\frac{1}{{\bf k}^2}
\left\{
1
+
\sum_{n=1}^{\infty}g_B^{2n}{\bf k}^{2n\epsilon}\frac{\tilde c_n(D)}{(4\pi)^{nD/2}}
\right\}
\,.
\label{momspacepot}
\ee
The coefficients $\tilde c_1(D)$ and $\tilde c_2(D)$ can be found 
in Ref.~\cite{Schroder:1999sg}. Throughout this paper we will use the notation 
$D\equiv 4+2\epsilon \equiv n+2\epsilon_n$, where $\epsilon_n=\frac{D-n}{2}$ parametrizes the (typically infinitesimal) difference to the closest integer dimension $n=4$, 3, 2.

After the Fourier transformation to position space Eq.~(\ref{momspacepot}) becomes
(see for instance Ref.~\cite{Pascual:1984zb})
\bea
\label{VsBD}
V_{s,B}&=&-C_Fg_B^2
\sum_{n=0}^{\infty}\frac{g_B^{2n}r^{-2(n+1)\epsilon}}{r}
\frac{\tilde c_{n}(D)}{(4\pi)^{nD/2}}
\frac{\Gamma(1/2+(n+1)\epsilon)}{2^{2-2n\epsilon}\pi^{3/2+\epsilon}\Gamma(1-n\epsilon)}
\\
\nn
&\equiv&
-C_Fg_B^2\sum_{n=0}^{\infty}\frac{g_B^{2n}c_n(D)r^{-2(n+1)\epsilon}}{r}\,.
\eea

The singlet static energy can be considered to be an observable for our purposes. 
It consists of the potential, which is a Wilson coefficient, and an ultrasoft 
contribution\footnote{If one has enough precision non-perturbative effects at the 
hadronization scale $\lQ \sim g^{\frac{2}{4-D}}$ should also be included. We will discuss 
them in Sec.~\ref{sec3D}.}, either bare or renormalized:
\be
E_s(r)=
V_{s,B}+\delta E^{us}_{s,B}
=
V_{s,\MS}+\delta E^{us}_{s,\MS}
\,.
\label{Estot}
\ee
The ultrasoft contribution can be expressed in a compact form at NLO in the multipole expansion (but exact to any order in $\al$) through the 
chromoelectric correlator. It reads (in the Euclidean)
\be
\delta E^{us}_{s,B}
  = V_A^2{T_F \over (D-1) N_c} {\bf r}^2 \int_0^\infty \!\! dt  e^{-t(V_{o,B}-V_{s,B})} 
  \langle vac|
g{\bf E}_E^a(t) 
\phi_{\rm adj}^{ab}(t,0) g{\bf E}_E^b(0) |vac \rangle
\label{deltaVUS}
\,.
\end{equation}

Concrete results for the ultrasoft corrections in $D$ dimensions are 
known at one loop since Ref.~\cite{Pineda:1997ie} (see also~\cite{short,KP1}). 
The pNRQCD one-loop computation yields
\be
\delta E^{us}_{s,B}({\rm 1-loop})
=
-g^2C_FV_A^2(1+\epsilon)\frac{\Gamma(2+\epsilon)\Gamma(-3-2\epsilon)}{\pi^{2+\epsilon}}
{\bf r}\,\Delta V^{3+2\epsilon}_B{\bf r}\,,
\label{USbare1loop}
\ee
where we have defined $\Delta V \equiv V_o-V_s$.
The two-loop bare expression can be deduced from 
the results obtained in Refs.~\cite{Eidemuller:1997bb,Brambilla:2006wp} and reads
\be
\label{USbare2loop}
\delta E^{us}_{s,B}({\rm 2-loop})
=
g^4C_FC_AV_A^2\Gamma(-3-4\epsilon)
\left[
{\cal D}^{(1)}(\epsilon)-(1+2\epsilon){\cal D}_1^{(1)}(\epsilon)
\right]
{\bf r}\,\Delta V_B^{3+4\epsilon}{\bf r}\,,
\ee
where
\be
{\cal D}^{(1)}(\epsilon)
=
\frac{1}{(2\pi)^2}\frac{1}{4\pi^{2+2\epsilon}}\Gamma^2(1+\epsilon)g(\epsilon)\,,
\ee
\be
{\cal D}_1^{(1)}(\epsilon)
=
\frac{1}{(2\pi)^2}\frac{1}{4\pi^{2+2\epsilon}}\Gamma^2(1+\epsilon)g_1(\epsilon)\,,
\ee
and
\be
g(\epsilon)=\frac{2 \epsilon^3+6 \epsilon^2+8 \epsilon+3}{\epsilon \left(2 \epsilon^2+5 \epsilon+3\right)}
-\frac{2 \epsilon \Gamma
   (-2 \epsilon-2) \Gamma (-2 \epsilon-1)}{(2 \epsilon+3) \Gamma (-4 \epsilon-3)}
\,,
\ee
\be
g_1(\epsilon)=\frac{6 \epsilon^3+17 \epsilon^2+18 \epsilon+6}{\epsilon^2 
\left(2 \epsilon^2+5 \epsilon+3\right)}+\frac{4 (\epsilon+1)
   n_f T_f}{\epsilon (2 \epsilon+3) N_c}+\frac{2
   \left(\epsilon^2+\epsilon+1\right) \Gamma (-2 \epsilon-2) \Gamma (-2 \epsilon-1)}{\epsilon (2 \epsilon+3)
   \Gamma (-4 \epsilon-3)}
\,.
\ee

\subsection{Renormalization, generalities}
\label{RenAndGen}

The bare parameters of the theory are $\al_B$ ($g_B$) and the potentials $V_B=V_{\{s,o,A,B\},B}$. In our convention 
$\al_B$ has integer mass dimension and is related to $g_B$ by
\be
\alpha_B=\frac{g_B^2\nu^{2\epsilon_n}}{4\pi}\,,
\ee
where $\nu$ is the renormalization scale.
It has a special status since 
it does not receive corrections from other Wilson coefficients of the 
effective theory. Therefore, 
it can be renormalized multiplicatively:
\be
\al_B=Z_{\al}\al
\,,
\ee
where
\be
Z_{\al}
=1+
\sum_{s=1}^{\infty}Z^{(s)}_{\al}\frac{1}{\epsilon_n^s}\,.
\ee
The renormalization group equation (RGE) of $\al$ is
\be
\nu\frac{d}{d\nu}\al\equiv \al\beta(\al;\epsilon_n)=2\epsilon_n\al+\al\beta(\al;0)\,.
\ee
In the limit $\epsilon_n \rightarrow 0$
\be
\nu\frac{d}{d\nu}\al\equiv \al\beta(\al;0)
\equiv \al\beta(\al)=-2\al \frac{d}{d\al}Z^{(1)}_{\al}\,.
\ee

The bare potentials $V_B$ in position space have integer mass dimensions (note that this is not true in 
momentum space) and, due to the structure of the theory, we do not renormalize them multiplicatively, 
see the discussion in Ref.~\cite{RG}. We define
\be
\label{VBsplitting}
V_B=V+\delta V\,.
\ee
$\delta V$ will generally depend on the (matching) coefficients of the effective theory, i.e. on $\al$ and $V$ 
and on the number of space-time dimensions.
In $D(n)$ dimensions, using the MS renormalization scheme, we define
\be
\delta V
=
\sum_{s=1}^{\infty}Z^{(s)}_{V}\frac{1}{\epsilon_n^s}\,.
\ee

From the scale independence of the bare potentials
\be
\nu \frac{d}{d\nu}V_B=0
\,,
\ee
one obtains the RGE's of the different
renormalized potentials. They can schematically be written as one (vector-like) equation including all potentials:
\be
\nu \frac{d}{d\nu}V=B(V)\,, \label{VRGE}
\ee
\be
B(V)\equiv -\left(\nu \frac{d}{d\nu}\delta V\right)\,.
\ee
Note that Eq.~(\ref{VRGE}) implies that all the $1/\epsilon_n$ poles disappear once the derivative 
with respect to the renormalization scale is performed.
 This imposes some constraints on $\delta V$:
\bea
\label{Z1}
{\cal O}(1/\epsilon_n): \qquad
&&B(V)=-2\al \frac{\partial}{\partial\al}Z^{(1)}_{V}\,,\; \label{1loopBV}
\\
{\cal O}(1/\epsilon^2_n): \qquad
&&
\label{Z2}
B(V)\frac{\partial}{\partial V}Z^{(1)}_{V}
+
\al \beta (\al)\frac{\partial}{\partial\al}Z^{(1)}_{V}
+
2\al \frac{\partial}{\partial\al}Z^{(2)}_{V}=0\,,
\eea
and so on.

\section{pNRQCD (D=3)}
\label{sec3D}

In three dimensions the purely gluonic sector is superrenormalizable. As a consequence the coupling constant is not renormalized:
\be
\al_B=\al
\,.
\label{alphabare}
\ee

At leading order in the multipole expansion the singlet field of the quark-antiquark system is free, i.e. 
it does not interact with gluons, and renormalization scale independent.
In other words its renormalization constant (${\rm S}_B= Z_s^{1/2}{\rm S}$) is
\be
\label{ZsLO}
Z_s=1+{\cal O}(r^2)
\,.
\ee
Similarly the singlet potential is not renormalized at $\ord(r^0)$:
\be
\delta V_s={\cal O}(r^2)
\,.
\ee
\begin{figure}
\includegraphics[width=0.25 \textwidth]{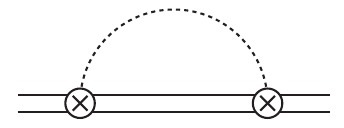}
\caption{\it One-loop contribution to the octet propagator. The dotted line 
represents the $A^0$ field.}
\label{usoctet3D}
\end{figure}

For the octet field the situation is different. Even at leading order in the multipole expansion it has 
a residual interaction with ultrasoft gluons. 
The octet potential receives an ultraviolet (UV) divergent correction from 
the one-loop self-energy diagram shown in Fig. \ref{usoctet3D}.
It gives
\be
Z^{(1)}_{V_o}
=
\frac{C_A}{2}\, \al + {\cal O}(r^2)
\,.
\label{ZVo}
\ee
Higher loop diagrams cannot contribute at $\ord(r^0)$.
This is because the coupling and/or the potentials must appear perturbatively (with positive powers) 
in the Z's. Since $\al$ has positive mass dimension, the potentials would appear with negative powers 
in higher loop corrections to Eq.~(\ref{ZVo}), which 
is not allowed by renormalizability. By the very same reason the octet field does not require renormalization at $\ord(r^0)$, 
i.e.
\be
Z_o=1 + \ord(r^2).
\ee
This will be all we need to know about the renormalization of the octet sector. 

The operators that appear at ${\cal O}(r)$ in the Lagrangian, $V_{A/B}$, 
are not renormalized at $\ord(r^0)$ either, i.e.
\be
\label{ZVAB}
Z_{A/B}=1+ \ord(r)
\,.
\ee
The reason is, as mentioned before, that the coupling and the potentials have to appear perturbatively in the counterterms. Otherwise the renormalizability of the theory at LO of the multipole expansion would be spoiled. 
Moreover, we can take $V_A=V_B=1$, as ${\cal O}(\al)$ soft corrections would be multiplied by factors of $r$ and would move us away from 
the precision of $V_s$ aimed for at this paper. Actually, from inspection of the possible diagrams that will contribute 
at the soft scale, we know that $V_{A/B}=1+{\cal O}(\al^2)$~\cite{Brambilla:2006wp}.

We now focus on the singlet and the 
renormalization of $V_s$ beyond ${\cal O}(r^0)$. 
The singlet potential is IR safe up to soft one-loop order. At two soft loops in dimensional regularization IR poles up to $\ord(1/\epsilon_3^3)$ appear~\cite{Schroder:1999sg}.
The ultrasoft computation in pNRQCD yields the following results for the counterterms 
\bea
\label{Z1Vs}
Z^{(1)}_{V_s,X}
&=&
-r^2 \Delta V_X^2\al C_F V_A^2\frac{1}{4}
-r^2 \Delta V_X \, C_A\al^2 C_F V_A^2\frac{1}{4}\
\\
\nn
&&
-r^2\al^3 C_FV_A^2
\frac{\left(13 \pi ^2-2208\right) C_A^2+8 \left(19\pi^2+144\right) 
C_A T_Fn_f-48 T_Fn_f \left(4 \left(\pi ^2-10\right) C_F+\pi ^2T_Fn_f\right)}{2304}\,,
\eea
\be
\label{Z2Vs}
Z^{(2)}_{V_s,X}
=
-r^2 \Delta V_X \al^2 C_FC_AV_A^2\frac{1}{8}
-r^2\al^3 C_FC_A^2V_A^2\frac{1}{24}\,,
\ee
\be
\label{Z3Vs}
Z^{(3)}_{V_s}
=
-r^2\al^3 C_FC_A^2V_A^2\frac{1}{48}\,,
\ee
\be
\label{ZnVs}
Z^{(n)}_{V_s}=0 \quad \forall \quad n>3\,.
\ee

These are the complete  ${\cal O}(r^2)$ results. We used the MS renormalization scheme to derive them 
and we checked that their form is invariant under scheme transformations that amount to a change 
of the renormalization scale: $\nu \rightarrow \nu\, c_X^{-1}$, where $c_X$ is a $\epsilon_3$-independent constant ($c_{\rm MS}=1$). We refer to this class of renormalization schemes as ``global'' and indicate globally renormalized quantities by an index ``$X$'' representing the scheme (X=MS, $\msb$, etc.) as e.g. $\Delta V_X$ in the above equations.

The fact that one can renormalize the 
potential with a finite number of terms at a given order in the multipole expansion (at ${\cal O}(r^2)$: $Z^{(n)}_{V_s}=0$ for $n>3$)
reflects the super-renormalizability of the theory. 
Let us now explain how Eqs.~(\ref{Z1Vs}-\ref{Z3Vs}) are obtained. 

1) The first term in Eq.~(\ref{Z1Vs}) comes from the $1/\epsilon_3$ divergence of the ultrasoft one-loop correction in Eq.~(\ref{USbare1loop}).
It has a scheme independent form and fixes, together with Eqs.~(\ref{Z1}) and~(\ref{Z2}), the first term of Eq.~(\ref{Z2Vs}) and 
Eq. (\ref{Z3Vs}). Of course it is also possible to compute these $1/\epsilon_3^2$ and $1/\epsilon_3^3$
divergences directly. They are generated by the entanglement of the ultrasoft one-loop diagram
with the octet self energy in Fig.~\ref{usoctet3D}.\footnote{
In fact one can even perform an exact resummation of these diagrams (at least of the divergent pieces) 
to handle all those divergences at once.} Other types of diagrams with the same number of loops are less divergent.  

2) The second term in Eq.~(\ref{Z1Vs}) follows from the remaining $1/\epsilon_3$ divergence in the ultrasoft two-loop computation, Eq.~(\ref{USbare2loop}), once all subdivergences (associated with the octet potential) have been subtracted. 
We have checked its (global) scheme independence explicitly.
This result combined with Eq.~(\ref{Z2}) then fixes the 
second term of Eq.~(\ref{Z2Vs}).

3) We now compare the results of 1) and 2) (with $V_A=1$) to the two-loop calculation of the soft contribution, 
which equals the bare static potential of Ref. \cite{Schroder:1999sg}. 
Keep in mind that the comparison is carried out in $D=3+2\epsilon_3$ dimensions and $\Delta V$ is a polynomial in $\epsilon_3$. 
We find perfect agreement of the soft IR $1/\epsilon_3^3$ and $1/\epsilon_3^2$ singularities and the corresponding ultrasoft counterterms in Eq.~(\ref{Z2Vs}) and~(\ref{Z3Vs}).\footnote{Actually, 
if we compare with Eq. (3.51) of Ref.~\cite{Schroder:1999sg}, we find a discrepancy, but if we 
explicitly sum up the expressions for the contributing diagrams given in the same reference the results agree. 
Most likely there is a typo in Eq. (3.51) of Ref.~\cite{Schroder:1999sg}.}
That means that the respective divergent parts of the bare quantities in Eq.~(\ref{Estot}) cancel.
Note that this can be understood as a non-trivial check of two independent determinations of these terms.

4) Our findings so far open the way to obtain the missing subleading UV divergence in Eq.~(\ref{Z1Vs}).
The key point is that the counterterms of the potential cannot contain negative powers of $\Delta V$. They would give rise to 
negative powers of $\ln (r \nu)$, which cannot be absorbed by the potential (cf. Eq.~(\ref{VsBD})). 
Hence $Z^{(1)}_{V_s}$ is a polynomial of $\Delta V$.
From the previous computations we know all terms of the polynomial except the constant at ${\cal O}(\Delta V^0)$. This term can be inferred 
from the (soft) result in Ref.~\cite{Schroder:1999sg}.
We simply subtract all the $1/\epsilon_3^3$ and $1/\epsilon_3^2$ counterterms as well as the first two $1/\epsilon_3$ terms according to Eq. (\ref{Z1Vs}). The remaining (IR) $1/\epsilon_3$ divergence must then match the last term in Eq. (\ref{Z1Vs}) in order to render Eq.~(\ref{Estot}) finite. 
In this way we fix the $\Delta V^0$ piece of $Z^{(1)}_{V_s}$ indirectly.
A consistency check of this computation is that the result is really independent of $\Delta V$ (i.e. free of $\ln(r)$) and invariant within the class of global renormalization schemes defined above.
Thus we have obtained the complete renormalization structure of $V_s$ at ${\cal O}(r^2)$.

\subsection{pNRQCD RG}

We will now derive the RG evolution of the matching coefficients of the effective theory. 
The running of the coupling constant is trivial, since it is not renormalized in three dimensions, 
as follows from Eq. (\ref{alphabare}):
\be
\nu\frac{d}{d\nu}\al=0 + \ord(\epsilon_3)
\,.
\ee
As a consequence of Eq.~(\ref{ZVAB}) $V_{A/B}$ does not run either at $\ord(r^0)$. 
Similarly there is no running of the singlet static potential at LO in the multipole expansion, whereas 
the running of the octet potential according to Eqs.~(\ref{1loopBV}) and~(\ref{ZVo}) reads
\be
\nu\frac{d}{d\nu}V_o=-C_A\al
\,.
\ee
We are interested in the running of the octet potential only insofar as it enters the running 
of $\Delta V$, which including the tree-level matching condition is then given by
\be
\label{VoRG}
\Delta V_X(r;\nu)= V_{o,X}(r;\nu)-V_{s,X}(r;\nu)  =-\al C_A\ln(r\nu d_X)+\ord(r^2)+\ord(\epsilon_3)
\,.
\ee
The index $X$ stands for the scheme (e.g. MS or $\MS$):
\be
d_{\rm MS}=e^{\gamma_E/2}\sqrt{\pi} \simeq 2.36546,
\qquad 
d_{\rm \MS}=d_{\rm MS}\, c_{\MS}^{-1}=e^{\gamma_E}/2 \simeq 0.890536,
\ee
with $c_{\MS}= e^{1/2(\ln (4\pi) - \gamma_E )}$.
This result will be sufficient for our purpose, which is the computation of the singlet static potential and energy. 

From the counterterms determined in the previous subsection, we can derive
the complete running of the singlet static potential at ${\cal O}(r^2)$:
\bea
\label{VsRGeq}
\nu\frac{d}{d\nu}V_{s,X} &=& \frac{C_F}{2}V_A^2r^2(\Delta V_X)^2\al+C_FV_A^2r^2 \Delta V_X C_A\al^2
\\
&&
+\,
r^2\al^3C_FV_A^2
\frac{(13 \pi ^2-2208) C_A^2+8 (19\pi^2+144) 
C_A n_f T_F-48 n_fT_F (4 (\pi ^2-10) C_F+\pi ^2n_f T_F)}{384}\,.
\nn
\eea
Its form is globally scheme independent. 

By solving the RG equations we obtain for the ${\cal O}(r^2)$ contribution
\be
\label{VsRG}
V_s^{\rm MS}(\nu)=V_s^{\rm MS}(r;\nu \!=\! \frac1{r})+V_s^{\rm RG;MS}(r;\nu)\,,
\ee
where
\begin{eqnarray}
 \lefteqn{V_s^{\rm RG;MS}(r;\nu)=C_F r^2 \alpha ^3 }  \nonumber\\
&\times&
 \Bigg\{\frac{1}{6} C_A^2 \ln^3(r \nu ) \nonumber\\
&&
+\frac{1}{4} C_A^2 \big[-2+\gamma_E +\ln \pi \big] \ln^2(r \nu ) \nonumber\\
&&
+\bigg[ C_A^2 \Big(\frac{13\pi ^2}{384}+\frac{1}{8} \big(\gamma_E ^2 -4 \gamma_E - 46 + (2 \gamma_E - 4)\ln \pi+\ln^2 \pi   \big) \Big) \nonumber\\
&&
+n_f T_F \Big(C_A \Big(3+\frac{19 \pi ^2}{48}\Big) +C_F \Big(5-\frac{\pi ^2}{2}\Big)\Big) - (n_f T_F)^2 \frac{ \pi ^2}{8}\bigg] \ln (r \nu)
\Bigg\}
\label{Vsrunning}
\end{eqnarray}
is the running and 
\begin{eqnarray}
  \lefteqn{ V_s^{\rm MS}(r;\nu \!=\! \frac1{r}) = 
  C_F \alpha \ln (r^2 \nu^2_s\pi e^{\gamma_E} )  +  \frac{\pi}{4} C_F (7 C_A-4 n_f T_F) \alpha^2 r
   } \nonumber\\
&+C_F r^2 \alpha ^3&
 \Bigg\{C_A^2 \bigg[\frac{\pi ^2}{2304} (39 \gamma_E-715 +564 \ln 2+39 \ln \pi ) + \frac{209}{24}-\frac{43}{24} \zeta (3) \nonumber\\
&&
 \qquad+\frac{\gamma_E^3 - 6 \gamma_E^2 - 138\gamma_E + 
 3(\gamma_E^2 - 4 \gamma_E -46) \ln \pi +3 (\gamma_E-2) \ln^2 \pi +\ln^3 \pi}{48}\bigg] \nonumber\\
&&
 + n_fT_F C_F \Big[\frac{\pi^2}{24}  (15-6 \gamma_E -8 \ln 2-6 \ln \pi )+\frac{1}{2} (5 \gamma_E-8 +5 \ln \pi )\Big] \nonumber\\
&&
+ n_fT_F C_A \Big[\frac{\pi^2}{288} (57 \gamma_E -97 +12 \ln 2+57 \ln \pi )+\frac{1}{6} (9 \gamma_E -22 +9 \ln \pi +10 \zeta (3))\Big] \nonumber\\
&&
+(n_f T_F)^2 \Big[\frac{\pi^2}{48}(7-3 \gamma_E -\ln (16 \pi ^3) ) \Big] \Bigg\}
\label{Vs1r}
\end{eqnarray}
is the initial matching condition. Note that the tree-level and one-loop matching conditions have been included.  
For the two-loop initial condition we used the bare data of Ref.~\cite{Schroder:1999sg}. 

At tree-level there is a dependence on another factorization scale $\nu_s \gg 1/r$. The logarithmic dependence of $V_s$ on $\nu_s$ would cancel the (hard) IR scale dependence of a large but finite mass $m$ ($m \gg \nu_s$). Once the mass is included, the static energy reads
\be
E_s(r)=2m(\nu_s)+V_s(r;\nu_s;\nu)+\delta E_s(\nu)
\,.
\ee
The $\nu_s$ dependence of $m$ is the three dimensional relic of the four dimensional pole mass renormalon. 
In this work we only care about the ultrasoft ($\nu$) scale dependence. Therefore, we do not need to explicitly consider $\nu_s$, which only appears in the initial matching condition, or, if combined with the mass, is replaced by $m$ (up to a constant).
On the other hand, note that for $\Delta V=V_o(\nu)-V_s(\nu)$ the dependence on $\nu_s$ disappears. 
This is analogous to the ultraviolet renormalon cancellation that takes place in four dimensions, but does not mean that 
$\Delta V$ is renormalon free (in 4D). There is a leftover infrared renormalon, which is reflected by the remaining $\nu$ dependence in three dimensions. 

Note that we have now obtained the exact $O(r^2)$ contribution to the static potential (i.e. the soft contribution to the 
static energy). There is nothing left. Moreover, by setting $\nu \sim \Delta V$ potentially large logarithms are resummed.

The above results have been presented in the MS scheme. It is possible to change to a different renormalization scheme. If we rewrite Eq.~(\ref{VsRG}) in terms of $\Delta V$ all the global scheme dependence gets encapsulated in the $\Delta V$'s and Eq. (\ref{Vsrunning}) becomes
\bea
\nn
&&
V_s^{\rm RG;MS}(r;\nu)
=
-\frac{C_F}{6C_A}V_A^2r^2\left(\Delta V^3_X(r;\nu)-\Delta V^3_X(r;1/r)\right)
-\frac{C_F}{2}V_A^2r^2\al\left(\Delta V^2_X(r;\nu)-\Delta V^2_X(r;1/r)\right)
\\
&&
\nn
-
r^2\al^2\frac{C_F}{C_A}V_A^2
\frac{\left(13 \pi ^2-2208\right) C_A^2+8 \left(19\pi^2+144\right) 
C_A n_fT_F-48 n_fT_F \left(4 \left(\pi ^2-10\right) C_F+\pi ^2 n_fT_F \right)}{384}
\\
   &&
   \qquad
   \times \left(\Delta V_X(r;\nu)-\Delta V_X(r;1/r)\right)
\,.
\label{VsRGMSinDeltaV}
\eea
In a general global renormalization scheme the $[\Delta V_X(r;1/r)]^n$ terms in Eq.~(\ref{VsRGMSinDeltaV}) vanish, when the respective matching condition is added, leaving a (globally) scheme independent constant.

\subsection{$\delta E^{us}_s$: ultrasoft effects up to NNLL order}

We proceed with computing the static singlet energy. We have already derived the RG improved 
expression for the potential in the previous subsection. Therefore, the computation that is left (if we neglect non-perturbative effects) 
addresses the ultrasoft contribution. The compact expression Eq.~(\ref{deltaVUS}) includes all ultrasoft effects at 
${\cal O}(r^2)$ and at any 
order in the $g^2/\Delta V$ expansion. At present concrete results are available at one, $\ord (g^2)$, 
and two loops, $\ord(g^4)$, given in Eqs.~(\ref{USbare1loop}, \ref{USbare2loop}). After minimal subtraction they read 
\be
\delta E^{us}_{s,\rm MS}(\text{1-loop})
=
\frac1{4} \al\, C_FV_A^2 r^2\, \Delta V_{\rm MS}^{2}
\left(
1+\gamma_E-\ln(4\pi)+2\ln \Big[ \frac{\Delta V_{\rm MS}}{\nu} \Big]
\right)
\ee
\bea
\nn
\delta E^{us}_{s,\rm MS}(\text{2-loops})
&=&
-\frac{1}{2} \al^2\, C_F \Delta V_{\rm MS} V_A^2 r^2  
\left(
C_A\ln^2\Big[ \frac{\Delta V_{\rm MS}}{\nu } \Big] 
+
C_A\ln \Big[ \frac{\Delta V_{\rm MS}}{\nu } \Big] 
\left(
\gamma_E-\ln(4\pi)
\right) 
   \right.
   \\
   &&
   \left.
   +C_A
   \big(
   6+\frac1{4}\left(\gamma_E-\ln(4\pi)\right)^2
   \big)
   -2n_fT_F
   \right)
   \label{deltaUS2loop}
\eea 
Setting $\nu \sim \Delta V_X(\nu)$ resums large logarithms in the potential and minimizes them in the ultrasoft contributions.
Expressed as a double expansion in $\al r$ and $1/\ln(r\Delta V)$
the result for the static energy reads with ${\cal O}(\al^2r^2)$ and NNLL accuracy
\bea
E_s(r)&=&
C_F \alpha \ln (r^2 \nu^2_s  \pi e^{\gamma_E})  +  \frac{\pi}{4} C_F (7 C_A-4 n_f T_F) \alpha^2 r+
V_s^{\rm RG;MS}(r;\nu\!=\!\Delta V)+\delta E^{us}_{s,\rm MS}(\nu\!=\!\Delta V) \nn\\
&=&
C_F \alpha \ln (r^2 \nu^2_s \pi e^{\gamma_E} 
)   +  \frac{\pi}{4} C_F (7 C_A-4 n_f T_F) \alpha^2 r  \nn
\\
&&+C_F \alpha ^3\, r^2
\Bigg\{
\frac{1}{6} C_A^2 \ln ^3(r \Delta V_{\rm MS}) 
+\frac{1}{4} C_A^2 (2 \gamma_E -1 - 2\ln 2 ) \ln ^2(r \Delta V_{\rm MS}) \nonumber\\
&&+\bigg[C_A^2 \Big(\frac{13 \pi ^2}{384}+\frac{1}{8} \left(4 \gamma_E ^2+4 \ln ^2 2-2\gamma_E  (1+ 4 \ln 2)-2 (11+\ln \pi )\right)\Big)\nonumber\\
&&
\hspace{2.5 ex} + n_f T_F \Big(C_A \big(2+\frac{19 \pi ^2}{48}\big) +C_F \big(5-\frac{\pi ^2}{2}\big)\Big)- (n_f T_F)^2 \frac{ \pi
   ^2}{8}\bigg] \ln (r \Delta V_{\rm MS}) \Bigg\} +{\cal O}(\al^3 r^2 \ln^0)
\,,
\label{NNLL}
\eea
where the omitted $\ord(\al^3 r^2)$ terms do not contain logarithms of $r \Delta V$.
Eq.~(\ref{NNLL}) is renormalization scale independent up to $\ord(\frac{\al^4 r^2}{\Delta V})$. 
Furthermore the explicit scheme dependence of $V_s^{\rm RG}$ and $\delta E^{us}_{s}$ and the 
implicit scheme dependence of Eq.~(\ref{NNLL}) through the logarithms of $\Delta V$ cancel up to $\ord(\al^3 r^2 \ln^0)$.

Finally we would like to note that the condition 
$\nu_{us} \equiv \Delta V_X(\nu_{us})$, produces a factorization-scale independent scale
that is non-perturbative in $\al$, 
\be
\nu_{us} \equiv \Delta V_X(\nu_{us})=C_A\al\, W(1/(C_A\al\, d_X r))\,.
\ee
$W(z)$ is the Lambert function and has the following expansion for 
large $|z|$ 
\be
\label{Lambert}
W(z)=\ln (z)-\ln (\ln (z))-\sum _{k=0}^{\infty }
   (-1)^k \ln ^{-k}(z) \sum _{j=1}^k \frac{\ln
   ^j(\ln (z)) S_k^{(-j+k+1)}}{j!}
\,,
\ee
where $S_k^{(-j+k+1)}$ is the Stirling number of the first kind\footnote{see e.g.: http://functions.wolfram.com/IntegerFunctions/StirlingS1}.  
Therefore, $\nu_{us}$ resums a certain class of logarithms\footnote{Something similar happens in four dimension when solving the Schr\"odinger equation 
with the lowest order Coulomb potential. In that case however one sets $\nu_s=m\,C_F\al(\nu_s)$.}.

\subsection{Subleading ultrasoft effects: ${\cal O}(\al^3r^2)$}

Let us now consider higher order corrections to the result obtained in Eq. (\ref{NNLL}). 
We have already stated that, from Eq. (\ref{VsRG}), the soft contribution to the static energy, i.e. the potential, is known completely 
at ${\cal O}(\al^3r^2)$. This is not the case for the ultrasoft contribution at that order.  
The general structure of $\delta E^{us}$ at ${\cal O}(r^2)$ is
\be
\delta E_s^{us}
=
C_F\al V_A^2r^2(\Delta V)^2\sum_{n=0}^{\infty}\left(\frac{C_A\al}{\Delta V}\right)^{\!\! n}\sum_{s=0}^{n+1}c_{n,s}\ln^s\Big[\frac{\Delta V}{\nu}\Big]
\,.
\label{USstruc}
\ee
Interestingly the RG contains important information on the higher order ultrasoft contributions, namely on the logarithmic terms. Since the singlet static energy $E_s$ is a physical observable and therefore
factorization/renormalization scale independent we find at $\ord(r^2)$ (neglecting non-perturbative effects for the moment)
\be
\nu\frac{d}{d\nu}  \delta E_s^{us} =-B(V_s)=-C_F\al V_A^2r^2\sum_{n=0}^2 B_n(C_A\al)^n(\Delta V)^{2-n}
\,,
\ee
where the $B_n$ coefficients can be read off Eq. (\ref{VsRGeq}). 
We obtain the following relation between the coefficients at different orders in $\Delta V$ (note that $\Delta V$ is scale dependent 
itself):
\be
\label{cnrelation}
B_n\delta_{s,0}+(n-3)c_{n-1,s}-(s+1)(c_{n-1,s+1}+c_{n,s+1})=0
\ee
with $c_{n,s}=0$ if $s > n+1$ or $n<0$ or $s<0$ and $B_n=0$ if $n>2$ or $n<0$.

In particular, Eq. (\ref{cnrelation}) allows to fix all logarithmic terms of the three-loop 
ultrasoft computation:
\begin{eqnarray}
 \lefteqn{\delta E_{s,\rm us}^{\rm 3loop;MS}(r;\nu)=C_F r^2 \alpha ^3 }  \nonumber\\
&\times&
 \Bigg\{\frac{1}{6} C_A^2 \ln ^3\Big[\frac{\Delta V_{\rm MS}}{\nu }\Big] \nonumber\\
&&
 + \frac{1}{4} C_A^2  \left(2 + \gamma_E  - \ln (4 \pi ) \right) \ln ^2\Big[\frac{\Delta V_{\rm MS}}{\nu }\Big] \nonumber\\
&&
+ \bigg[
C_A^2  \Big(\frac{13 \pi ^2}{384}-\frac{11}{4}
+\frac{1}{2}\left(\gamma_E-\ln(4\pi)\right)+\frac{1}{8}\left(\gamma_E-\ln(4\pi)\right)^2
\Big)\nonumber\\
&&
+n_f T_F \Big(\big(2+\frac{19 \pi ^2}{48}\big) C_A+ C_F \big(5-\frac{\pi ^2}{2}\big)\Big)
-(n_f T_F)^2\frac{ \pi ^2}{8}\bigg] \ln\Big[ \frac{\Delta V_{\rm MS}}{\nu } \Big]
 +C_A^2c_{2,0}^{\rm MS}\Bigg\}
\,.
\label{US3loop}
\end{eqnarray}

Note that, though we cannot fix the $\nu$ independent constant $c_{2,0}$ from RG arguments, it can be computed from perturbation theory, 
but requires a three-loop pNRQCD computation which has not been performed yet.
The constant term of the static singlet energy at ${\cal O}(\al^3r^2)$, which is the only missing 
term in Eq.~(\ref{NNLL}) to reach N$^3$LL precision, 
is then given by 
\be
\label{NNNLLmatching}
V_s^{\rm MS}(r;\nu \!=\! \frac1{r})\Big|_{{\cal O}(\al^3)} + C_F C_A^2r^2 \al^3 c_{2,0}^{\rm MS}
\,.
\ee

\subsection{Non-perturbative effects: ${\cal O}(\al^4r^2/\Delta V)$}
\label{npsubsec}

At even higher orders in the $\al/\Delta V$ expansion, non-perturbative effects start to contribute. 
In order to study these effects related to loop momenta $k \sim \al$, we integrate out the $\Delta V$ scale. 
This means integrating out the octet field and ultrasoft gluons. The degrees of freedom left are the 
singlet field and non-perturbative gluons with energy and momentum of order $\al$. 
The resulting Lagrangian, including the leading order non-perturbative effects at ${\cal O}(r^2)$, reads
\begin{eqnarray}
 {\cal L}_{\rm np} =
{\rm Tr} \Biggl\{ {\rm S}^\dagger \left( i\partial_0  - V_s(r) -\delta E_s^{us}  \right) {\rm S} \Biggr\}
-  \frac{C_{np}}{\Delta V} {\rm Tr} \left\{  {\rm S}^\dagger {(g\bf E \cdot r})^2 \,{\rm S}\right\}
\label{pnrqcdnp}
\end{eqnarray}
for the case without light fermions ($n_f=0$). If we were to include light fermions there would also be operators of the form
\be
\delta {\cal L} \sim \frac{r^2}{\Delta V} {\rm Tr} \left\{  {\rm S}^\dagger \al^2\bar q q \,{\rm S}\right\}\,.
\ee
They could generate corrections to the static energy, due to the quark condensate, which are of the same parametric order as the 
purely gluonic ones. In the following we restrict ourselves to the purely gluonic case ($n_f=0$).

\begin{figure}
\includegraphics[width=0.6 \textwidth]{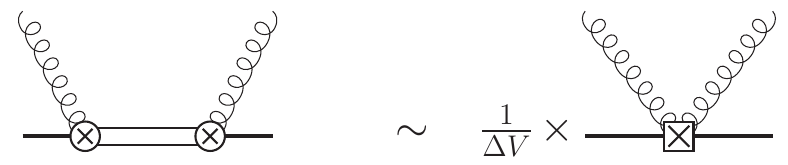}
\caption{\it Schematic tree-level matching procedure for the non-perturbative interaction operator in Eq.~(\ref{pnrqcdnp}). The pNRQCD octet field in the left diagram has been integrated out on the right hand side.}
\label{NonPertMatch}
\end{figure}

The coefficient of the non-perturbative operator in Eq.~(\ref{pnrqcdnp}) is $C_{np}=1$ at leading order in the 
$\frac{\al}{\Delta V}$ expansion. This result is obtained  
by matching to a pNRQCD tree-level diagram, where two gluons couple to the singlet field, as sketched in Fig.~\ref{NonPertMatch}.

The interaction with non-perturbative gluons produces a shift of the energy which is proportional to the gluon 
condensate in three dimensions\footnote{This is the first term of a Taylor series that can be easily obtained by expanding 
 Eq. (\ref{deltaVUS}) in $1/\Delta V$, see also Ref. \cite{Pineda:1996uk}:
\be
\delta E^{np}=\sum_{i=0}^{\infty}C_iO_i \,,
\label{higherNPcorr}
\ee
where
\be
C_i=(-1)^{2i+1}\frac{r^2}{(\Delta V)^{2i+1}} 
\ee
and ($H_g$ represents the gluonic Hamiltonian)
\be
O_i =
{T_F g^2 \over N_c(D-1)}
\langle vac \vert E^{a}_j
H_g^{2i}
E^{a}_j \vert  vac \rangle\,.
\ee
Using the equation of motion, the gauge fixing and Lorentz
covariance we obtain
\bea
&& O_i =-
{T_F g^2 \over N_c^2(D-1)}
v^{\beta_0}...v^{\beta_i}v^{\alpha_0}...v^{\alpha_i}
\\
\nonumber
&&
\times
\langle 0 \vert
{\rm Tr} \left(
[D_{\beta_1}(0),[...[D_{\beta_i}(0),G_{\beta_0 \rho}(0)]...]
[D_{\alpha_1}(0),[...[D_{\alpha_i}(0),{G_{\alpha_0}}^{\rho}(0)]...]
\right) \vert  0 \rangle
\,,
\eea
where $v$ is the velocity of the center of mass frame with $v^2=1$
(in the comoving frame $v=(1, {\bf 0})$) and the trace is taken in the
adjoint representation. Generally, however, Eq. (\ref{higherNPcorr}) does not cover all possible non-perturbative corrections at ${\cal O}(r^2)$.
} 
\be
\delta E^{np}_{s,B}=C^B_0O^B_0=\frac{r^2}{\Delta V_B}\frac{2\pi}{N_c(D-1)D}\langle \al\, G_{\mu \nu}^a G^{\mu \nu,a} \rangle_B\,.
\label{deltaEnp}
\ee
The leading ultraviolet divergence of the gluon condensate has been calculated in perturbation theory at 
four loops \cite{Schroder:2003uw}. The determination of the finite piece requires lattice simulations \cite{Hietanen:2004ew,Hietanen:2006rc}
and a computation to change from the lattice to dimensional regularization \cite{DiRenzo:2006nh}. 
Taking the result (in the Euclidean) from the last reference we have 
\bea
\lefteqn{\langle \al\, G_{\mu \nu}^a G^{\mu \nu,a} \rangle_B = \langle \al\, G_{\mu \nu}^a G^{\mu \nu,a} \rangle_{B,\rm Euclidean} =}&& \nn\\  
&&  6 \frac{\al^4}{\pi} C_F C_A^4 
 \biggl[
  \biggl( \frac{43}{12} - \frac{157}{768} \pi^2 \biggr)
 \biggl( -\frac{1}{8\epsilon_3}\bigg(\frac{e^{\gamma_E}}{4\pi}\bigg)^{\!\!4\epsilon_3}+
 \ln\frac{\nu}{2 C_A g^2} -\fr13
 \biggr)
 + \bG 
 + \ord(\epsilon_3)
 \biggr]
 \,,
 \label{eq:condensate}
\eea
\bea
B_G^{(SU(3))}=-0.2\pm 0.4({\rm MC}) \pm 0.4(\rm NSPT)\,. \nn
\eea

Eq.~(\ref{deltaEnp}) is the analog to the Voloshin-Leutwyler correction to 
the static potential in four dimensions \cite{Flory:1982qx,Bertlmann:1983pf}, which scales like $\frac{r^2}{\Delta V}\langle \al G^2 \rangle 
\sim r^3\langle \al G^2 \rangle$. In three dimensions we have instead $\frac{r^2}{\Delta V}\langle \al G^2 \rangle 
\sim \frac{r^2}{\ln r}\langle G^2 \rangle$.

Renormalizing the bare expression in Eq.~(\ref{deltaEnp}) in the $\rm MS$ scheme yields
\begin{eqnarray}
\delta E_{s, \rm MS}^{np}(\nu) 
=\frac{C_A^3 C_F r^2 \al^4 }{\Delta V_{\rm MS}} \bigg[
\Big(\frac{43}{6} - \frac{157}{384} \pi^2 \Big) \Big( \ln \Big[\frac{\nu}{C_A \al}\Big] - \frac12 (\ln (16 \pi)+ \gamma_E) - \frac{1}{8}\Big) + 2 B_G
\bigg]\,.
\label{NPlog}
\end{eqnarray}

To be consistent with our power counting this result has to be combined with the four-loop ultrasoft computation. Again the logarithm structure can be determined by 
demanding scale independence of $E_s(r)$, which results in the following renormalization group equation for the ultrasoft part
\be
\nu\frac{d}{d\nu}\delta E^{us}_s=-B(V_s)-\nu\frac{d}{d\nu}\delta E^{np}\,.
\ee
This equation allows us to fix the coefficients of the four-loop ultrasoft contribution:
\be
c_{3,4}=0\,,\qquad c_{3,3}=-c_{2,3}=-\frac{1}{6}\,,\qquad 
c_{3,2}=-c_{2,2}=-\frac{1}{2}-\frac{1}{4}\left(\gamma_E-\ln(4\pi)\right)\,, \qquad c_{3,1}=c_{3,1}^{UV}+c^{np}_{3,1}
\,.
\ee 
$c_{3,1}^{UV}=-c_{2,1}$ represents the coefficient of the ultraviolet logarithm and follows from Eq. (\ref{cnrelation}).
\be
c_{3,1}^{np}= \frac{43}{6} - \frac{157}{384} \pi^2
\ee
is associated with the infrared logarithm in the ultrasoft contribution, that cancels the renormalization scale dependence from the non-perturbative logarithm in Eq.~(\ref{NPlog}).

Thus in order to obtain the complete solution for the singlet static energy up to ${\cal O}(r^2\al^4/\Delta V)$, the missing pieces are purely 
perturbative: $c_{2,0}$ and $c_{3,0}$, i.e. the non-logarithmic three- and four-loop ultrasoft contributions.

The results given in Eqs.~(\ref{NNLL}),~(\ref{US3loop}) and~(\ref{NPlog}) 
have been written as an expansion in $1/\ln(r \Delta V)$. Alternatively one can use 
$1/\ln(C_A r\al)$ as the expansion parameter when setting $\nu_{us}=\Delta V_X$ and 
using Eq.~(\ref{Lambert}) to expand $\Delta V_X$ in powers of $(\ln(\ln(r\al)))^{n \le s} /\ln^s(r\al)$. 
This expansion might make sense 
as soon as non-perturbative terms like Eq.~(\ref{NPlog}) are included, since they can be organized in the same series.
Up to $\ord(r^2 \al^2)$ in the multipole expansion and $\rm N^3$LL order perturbation theory the singlet static energy then reads
\begin{eqnarray}
E_s(r)&=&    
C_F \alpha \ln (r^2 \nu^2_s \pi e^{\gamma_E})   +  \frac{\pi}{4} C_F (7 C_A-4 n_f T_F) \alpha^2 r  \label{FinalNNNLL}\\
&+& C_A^2 C_F r^2 \al^3 \Bigg\{ \frac{1}{6} \ln^3(C_A r \al ) \nn\\
&+& \ln^2(C_A r \al )
\bigg[
\frac{1}{2}\ln[-\ln(C_A r \al )]+\frac{1}{4} (2 \gamma_E - 2\ln 2 -1)
\bigg] \nn\\
&+& \ln(C_A r \al )
\bigg[
\frac{1}{2}\ln^2[-\ln(C_A r \al )]+(\gamma_E -\ln 2) \ln[-\ln(C_A r \al )]+ K
\bigg] \nn\\
&+&
\frac{1}{6}\ln^3[-\ln(C_A r \al )]+\frac{1}{2} (\gamma_E -\ln 2 +1) \ln^2[-\ln(C_A r \al )]+(K+\gamma_E -\ln 2) \ln[-\ln(C_A r \al )] \nn\\
&&+\frac{n_f T_F  \big(\pi ^2  (\frac{\ln 2}{8}-\frac{97}{96} )-11+5 \zeta(3) \big) }{3C_A} 
+\frac{\frac{\pi ^2}{48}  (7-4 \ln 2) (n_f T_F)^2 - C_F \left(4 - \pi ^2 \left(\frac{5}{8}-\frac{\ln 2}{3}\right)\right)
   n_f T_F}{C_A^2} \nn\\
&&-\frac{X^3}{6}-\frac{X^2}{2}  (\gamma_E -\ln 2 + 1) - X (K+\gamma_E - \ln 2) 
 +\frac{209 - \pi ^2 (\frac{715}{96} - \frac{47 \ln 2}{8})-43 \zeta (3)}{24} +c^X_{2,0} \nn
%
\Bigg\}
\end{eqnarray}
for a general renormalization scheme with $c_X=e^{\frac12(\gamma_E + \ln \pi) + X}$. 
\begin{eqnarray}
\lefteqn{K = B_2 - c_{1,0}^{\rm MS} + \frac{1}{8} (\gamma_E +\ln \pi ) (3 \gamma_E -\ln (16 \pi )) =}&&  \\
&&=\frac{(2+\frac{19 \pi ^2}{48}) n_f T_F}{C_A} - \frac{\frac{1}{8} \pi ^2 (n_f T_F)^2 + \frac{1}{2} C_F (\pi ^2-10) n_f T_F}{C_A^2}
+ \frac12 (\gamma_E -\ln 2)^2-\frac{11}{4} + \frac{13 \pi ^2}{384} \nn
\end{eqnarray}
is a manifestly scheme independent number. Thus the yet unknown ultrasoft constant $c_{2,0}$ defined by Eq.~(\ref{USstruc}) is expected to absorb the remaining scheme ($X$) dependence.

For $n_f=0$ we can even go one step further in the $1/\ln(C_A r\al)$ expansion and include the leading non-perturbative (Eq.~(\ref{NPlog})) 
as well as the N$^4$LL (ultrasoft four-loop) contribution to the static singlet energy by adding
\begin{eqnarray}
\delta E_s|_{\rm N^4LL}=
\lefteqn{\frac{C_A^2 C_F r^2 \al^3 }{\ln(C_A r \al )} \bigg[
\frac{1}{6}\ln^3[-\ln(C_A r \al )]+\frac{1}{2} (\gamma_E -\ln 2+1) \ln^2[-\ln(C_A r \al )] } && \nn\\
&& +\Big(K+\gamma_E -\ln 2 -\frac{43}{6} + \frac{157 \pi ^2}{384}\Big) \ln[-\ln(C_A r \al )] 
 + \Big(\frac{43}{6}-\frac{157 \pi ^2}{384}\Big) \big(\gamma_E +\ln (4 \pi )+\frac{1}{8}\big) \nn\\
&& -\frac{X^3}{6} - \frac{1}{2} X^2 (\gamma_E -\ln 2 +1) - X \Big(K+\gamma_E -\ln 2 - \frac{43}{6} +\frac{157 \pi^2}{384} \Big) - 2 B_G- c^X_{3,0}
\bigg]
\label{Final4loopNP}
\end{eqnarray}
to Eq.~(\ref{FinalNNNLL}). The $X$ dependent terms must be subtracted by corresponding terms in the ultrasoft four-loop constant $c_{3,0}$.

Another convenient choice for the renormalization scale is $\nu = C_A \al$, thus keeping the ultrasoft logarithms, but eliminating the logarithms in the non-perturbative finite parts. Eqs.~(\ref{FinalNNNLL}) and (\ref{Final4loopNP}) as parts of an expansion up to $\ord( \ln^{-1}(C_A r \al))$ remain of course unchanged.

\subsection{Comparison with lattice data}

The aim of this subsection is, on the one hand, to test how well the short-distance 3D lattice data can be reproduced by our theoretical 
expressions for the static singlet energy and, on the other hand, to try to extract numerical values for the ultrasoft 
three- and four-loop constants $c_{2,0}$ and $c_{3,0}$ from fits to this data.
Our final, most precise, theoretical expression for the static energy (for $n_f=0$) reads (combining Eqs. (\ref{NNLL}), (\ref{NNNLLmatching}) and (\ref{NPlog}))
\bea
\nn
E_s(r)&=&
C_F \alpha \ln (r^2 \nu^2_s \pi e^{\gamma_E} 
)   +  \frac{7\pi}{4} C_F C_A \alpha^2 r  
+C_A^2C_F \alpha ^3\, r^2
\Bigg\{
\frac{1}{6}  \ln ^3(r \Delta V_{\rm MS}) 
\\
&+&\ln ^2(r \Delta V_{\rm MS})\frac{1}{4}  (2 \gamma_E -1 - 2\ln 2 )  \nonumber\\
&+& \ln (r \Delta V_{\rm MS}) 
\bigg[\frac{13 \pi ^2}{384}+\frac{1}{8} \left(4 \gamma_E ^2+4 \ln ^2 2-2\gamma_E  (1+ 4 \ln 2)-2 (11+\ln \pi )\right)
 \bigg]
 \nonumber\\
&+& \bigg[ c_{2,0}^{\rm MS}
+ \frac{\pi ^2}{2304} (39 \gamma_E-715 +564 \ln 2+39 \ln \pi ) + \frac{209}{24}-\frac{43}{24} \zeta(3) 
\nn \\
&&+ \frac{1}{48} (\gamma_E +\ln\pi) \big(\ln^2\pi + 3(2 \ln2-1) \ln\pi -\gamma_E
    (3+18 \ln2 +4 \ln\pi )+7 \gamma_E^2+12 \ln^2 2 -66\big)
\bigg]
\nonumber\\
 &+&\frac{C_A \al }{\Delta V_{\rm MS}} \bigg[
\Big(\frac{43}{6} - \frac{157}{384} \pi^2 \Big) \Big( \ln \Big[\frac{\Delta V_{\rm MS}}{C_A \al}\Big] - \frac12 (\ln (16 \pi)+ \gamma_E) - \frac{1}{8}\Big) + 2 B_G
+ c_{3,0}^{\rm MS}
\bigg]
 \Bigg\} 
\,,
\label{Afinal}
\eea
or, alternatively (combining Eqs. (\ref{FinalNNNLL}) and (\ref{Final4loopNP})),
\begin{eqnarray}
E_s(r)&=&    
C_F \alpha \ln (r^2 \nu^2_s \pi e^{\gamma_E})   +  \frac{7\pi}{4} C_F C_A \alpha^2 r
+ C_A^2 C_F r^2 \al^3 \Bigg\{ \frac{1}{6} \ln^3(C_A r \al ) 
\nn\\
&+& \ln^2(C_A r \al )
\bigg[
\frac{1}{2}\ln[-\ln(C_A r \al )]+\frac{1}{4} (2 \gamma_E - 2\ln 2 -1)
\bigg] \nn\\
&+& \ln(C_A r \al )
\bigg[
\frac{1}{2}\ln^2[-\ln(C_A r \al )]+(\gamma_E -\ln 2) \ln[-\ln(C_A r \al )]+ K
\bigg] \nn\\
&+&
\bigg[
\frac{1}{6}\ln^3[-\ln(C_A r \al )]+\frac{1}{2} (\gamma_E -\ln 2 +1) \ln^2[-\ln(C_A r \al )]+(K+\gamma_E -\ln 2) \ln[-\ln(C_A r \al )] \nn\\
&&-\frac{X^3}{6}-\frac{X^2}{2}  (\gamma_E -\ln 2 + 1) - X (K+\gamma_E - \ln 2) 
 +\frac{209 - \pi ^2 (\frac{715}{96} - \frac{47 \ln 2}{8})-43 \zeta (3)}{24} +c^X_{2,0} \nn
\bigg]
\\
&+&
\frac{1}{\ln(C_A r \al )} \bigg[
\frac{1}{6}\ln^3[-\ln(C_A r \al )]+\frac{1}{2} (\gamma_E -\ln 2+1) \ln^2[-\ln(C_A r \al )]   \nn\\
&& +\Big(K+\gamma_E -\ln 2 -\frac{43}{6} + \frac{157 \pi ^2}{384}\Big) \ln[-\ln(C_A r \al )] 
+ \Big(\frac{43}{6}-\frac{157 \pi ^2}{384}\Big) \big(\gamma_E +\ln (4 \pi )+\frac{1}{8}\big) \nn\\
&& -\frac{X^3}{6} - \frac{1}{2} X^2 (\gamma_E -\ln 2 +1) - X \Big(K+\gamma_E -\ln 2 - \frac{43}{6} +\frac{157 \pi^2}{384} \Big) - 2 B_G- c^X_{3,0}
\bigg]
\Bigg\}\,.
\label{Bfinal}
\end{eqnarray}
The result is organized as a double expansion in powers of $\al r$ (multipole expansion) and 
$\al/\Delta V$ (ultrasoft expansion). We label the different terms of the multipole expansion as LO ($\sim \al \ln$), NLO ($\sim \al^2r$) and NNLO ($\sim \al^3r^2$). 
At NNLO in the multipole expansion the ultrasoft expansion sets in, and we label its different orders as LL [$\sim\! \ln^3 (r\Delta V) \!\sim\! \ln^3 (C_A r \al)$], NLL 
[$\sim\! \ln^2 (r\Delta V) \!\sim\! \ln^2\! (C_A r \al)$], \ldots, N$^n$LL [$\sim\! \ln^{3-n} (r\Delta V) \!\sim\! \ln^{3-n} (C_A r \al)$]. Note also 
that $\Delta V/(C_A\al) \sim  \ln r\Delta V$. 

\begin{figure}[t]
\includegraphics[width=0.7 \textwidth]{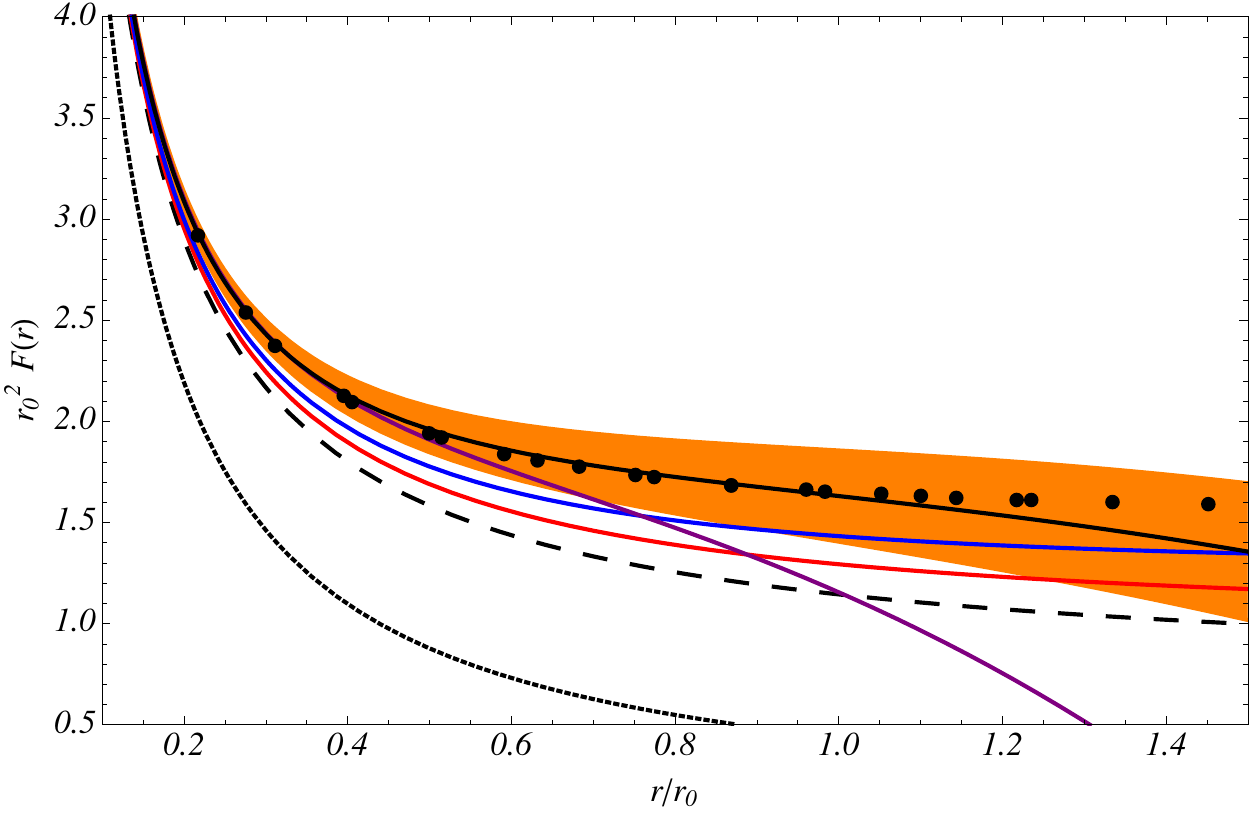}
\put(-180,190){\large a)\quad \bf SU(2)}
\put(-250,45){\sf LO}
\put(-200,60){\sf NLO}
\put(0,55){\sf \color{red} NNLL, \rm MS}
\put(0,66){\sf \color{blue} NNLL, $\msb$}
\put(-57,34){\sf \color{violet} N$^3$LL$+$np.\,log,\,\rm MS}
\put(-146,90){\sf N$^3$LL$+$np.\,log,\,\rm $\msb$}

\includegraphics[width=0.7 \textwidth]{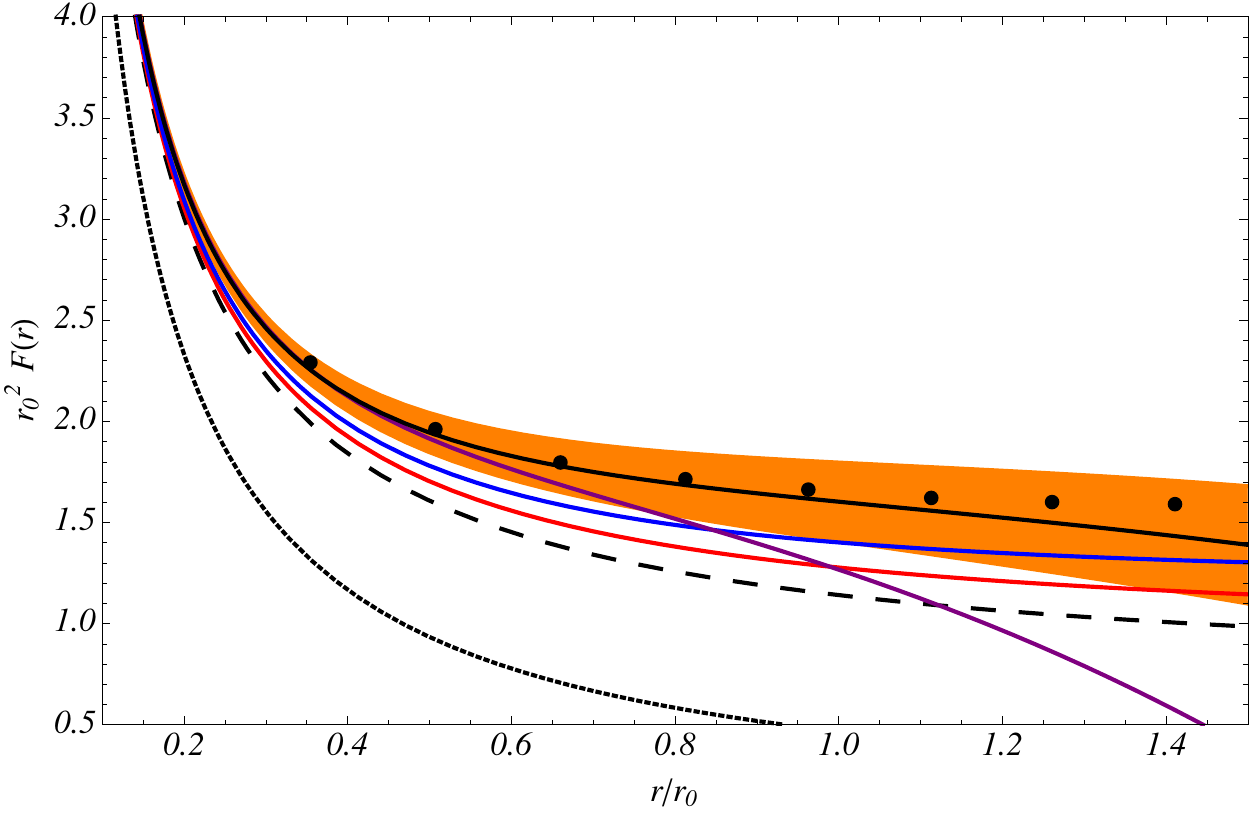}
\put(-180,190){\large b)\quad \bf SU(3)}
\put(-250,50){\sf LO}
\put(-200,61){\sf NLO}
\put(0,54){\sf \color{red} NNLL,\rm MS}
\put(0,64){\sf \color{blue} NNLL, $\msb$}
\put(-32,35){\sf \color{violet} N$^3$LL$+$np.\,log,\,\rm MS}
\put(-140,88){\sf N$^3$LL$+$np.\,log,\,\rm $\msb$}

\caption{\it Plots of the analytic results for the force $F(r)=dE_s(r)/dr$ in $r_0$ units with $n_f=0$ at different orders in the multipole and ultrasoft expansion in comparison to the available lattice data (black dots) for  $N_c=2$ (panel~a) and $N_c=3$ (panel~b). We show the LO (dotted) and NLO (dashed) curves in the multipole expansion. We also show some NNLO curves, which we label as 
 NNLL and $N^3$LL+np.\,log according to the order in the $\al/\Delta V$ expansion (see the main text). We have plotted our results in the $\rm MS$ and the $\msb$ scheme in order to make the scheme dependence visible and set $\nu=\Delta V_X$. The (orange) band represents the error of the $N^3$LL+np.\,log $\msb$ result from $c_{2,0}$ in Eq.~(\ref{c20fit}).
}
\label{Plots}
\end{figure}
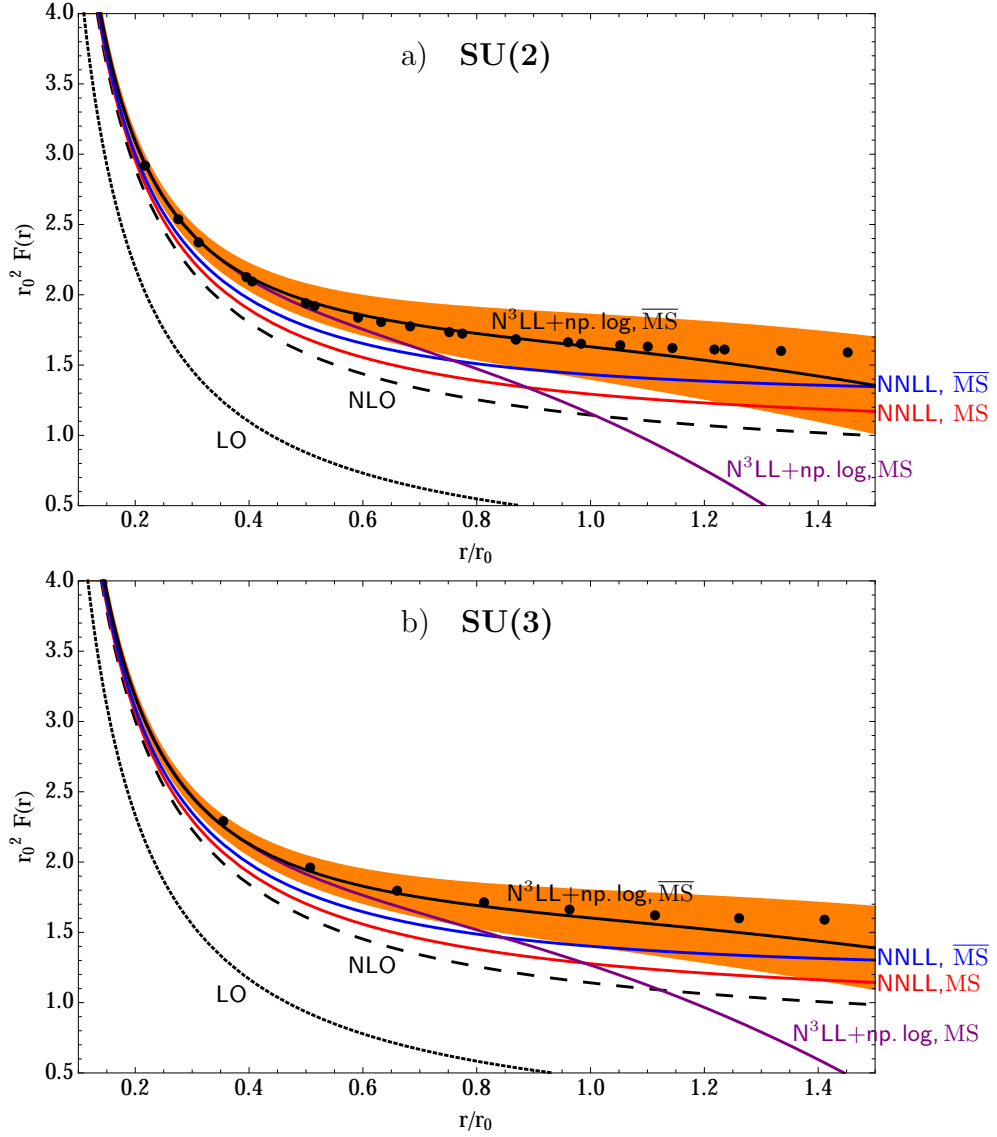

Eq.~(\ref{Afinal}) is written as an expansion 
in $1/\ln(r \Delta V)$, whereas Eq.~(\ref{Bfinal}) is written as an expansion in
$1/\ln(C_A r\al)$. The latter is obtained from Eq.~(\ref{Afinal}) 
using Eq.~(\ref{Lambert}) to expand $\Delta V_X$ in powers of $(\ln(\ln(r\al)))^{n \le s} /\ln^s(r\al)$. 
Eq.~(\ref{Bfinal}) is a strictly scheme independent N$^4$LL expression.
In other words, the scheme dependence of $c^X_{2,0}$ and $c^X_{3,0}$ cancels the explicit $X$ dependence in Eq.~(\ref{Bfinal}). 
Eq.~(\ref{Afinal}) includes an extra resummation of logarithms through Eq.~(\ref{Lambert}). 
We expect this resummation to improve the precision of the result but it also introduces a residual scheme dependence beyond N$^4$LL. 
Eq.~(\ref{Afinal}) is written in the MS scheme. 
The scheme dependence is entirely encapsulated in $\Delta V_{\rm MS}$ and the coefficients $c^{\rm MS}_{2,0}$ and $c^{\rm MS}_{3,0}$ so that a transformation to other schemes such as $\msb$ is straightforward.
Generally we expect the $\msb$ scheme to yield a better convergence since $\Delta V_{\msb}>\Delta V_{\rm MS}$.
Hence, we will use the $\msb$ expression as our default for the comparison with lattice and use the MS scheme and Eq.~(\ref{Afinal}) to analyze the scheme dependence of our results. 

The authors of Refs.~\cite{HariDass:2007tx,Luscher:2002qv,Meyer:2006gm} performed quenched ($n_f=0$) three dimensional lattice simulations 
of the force $F(r)=dE_s(r)/dr$ between two static color sources in the fundamental representation of SU($N_c$), which form a color singlet.
In Ref.~\cite{HariDass:2007tx} the number of colors is $N_c=2$ and in Ref.~\cite{Luscher:2002qv} $N_c=3$.\footnote{We don't 
consider the lattice data for $N_c=5$ in Ref.~\cite{Meyer:2006gm}, because we focus on short distances.} Unlike the static energy, the 
force does not contain the logarithmic (UV) divergence of the LO static potential, which depends on the regularization scheme. 
It is therefore the preferred quantity to study on the lattice.

\begin{figure}[t]
\includegraphics[width=0.7 \textwidth]{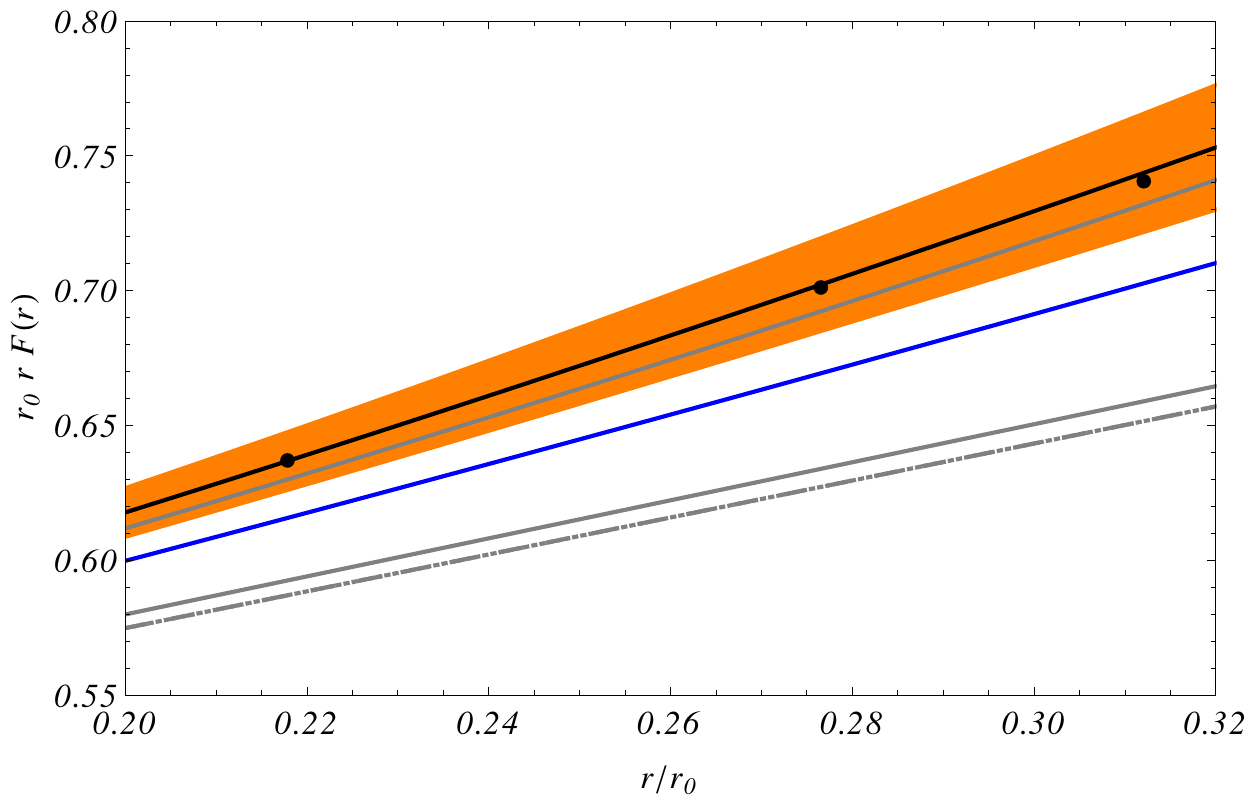}
\put(-195,185){\large a)\quad {\bf SU(2)}, $\msb$}
\put(-220,70){\sf\color{darkgray} NLO}
\put(-170,62){\sf\color{darkgray} LL (dotted)}
\put(-115,73){\sf\color{darkgray} NLL (dashed)}
\put(-80,114){\sf \color{blue} NNLL}
\put(-50,141){\sf\color{darkgray} N$^3$LL}
\put(-148,144){\sf N$^3$LL$+$np.\,log}

\includegraphics[width=0.7 \textwidth]{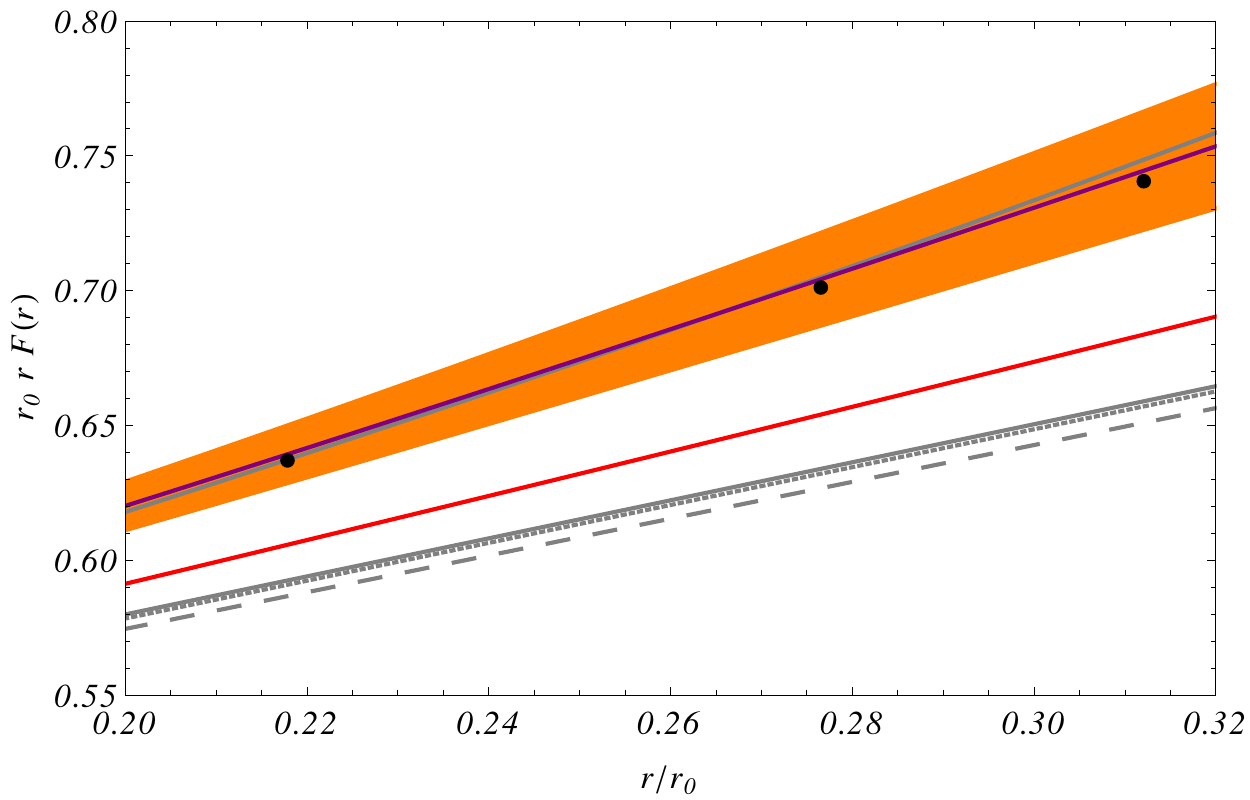}
\put(-195,185){\large b)\quad {\bf SU(2)}, MS}
\put(-40,108){\sf\color{darkgray} NLO}
\put(-170,62){\sf\color{darkgray} LL (dotted)}
\put(-115,73){\sf\color{darkgray} NLL (dashed)}
\put(-98,112){\sf \color{red} NNLL}
\put(-40,173){\sf\color{darkgray} N$^3$LL}
\put(-280,101){\sf\color{violet} N$^3$LL$+$np.\,log}

\caption{\it Plots of the analytic NNLO results for $r\,F(r)$ in $r_0$ units with $n_f=0$ in comparison to the available lattice data (black dots):
a) $\msb$ results for $N_c=2$ evaluated through different orders in the ultrasoft expansion: LL, NLL, NNLL, N$^3$LL and $N^3$LL+np.\,log (see text for details). The (orange) band again reflects the uncertainty of $c_{2,0}^\msb$ according to Eq.~(\ref{c20fit}).
b) Same curves as in panel~a) but in the {\rm MS} scheme.
In both plots we also show the NLO curve for comparison.
}
\label{Plotsshort}
\end{figure}
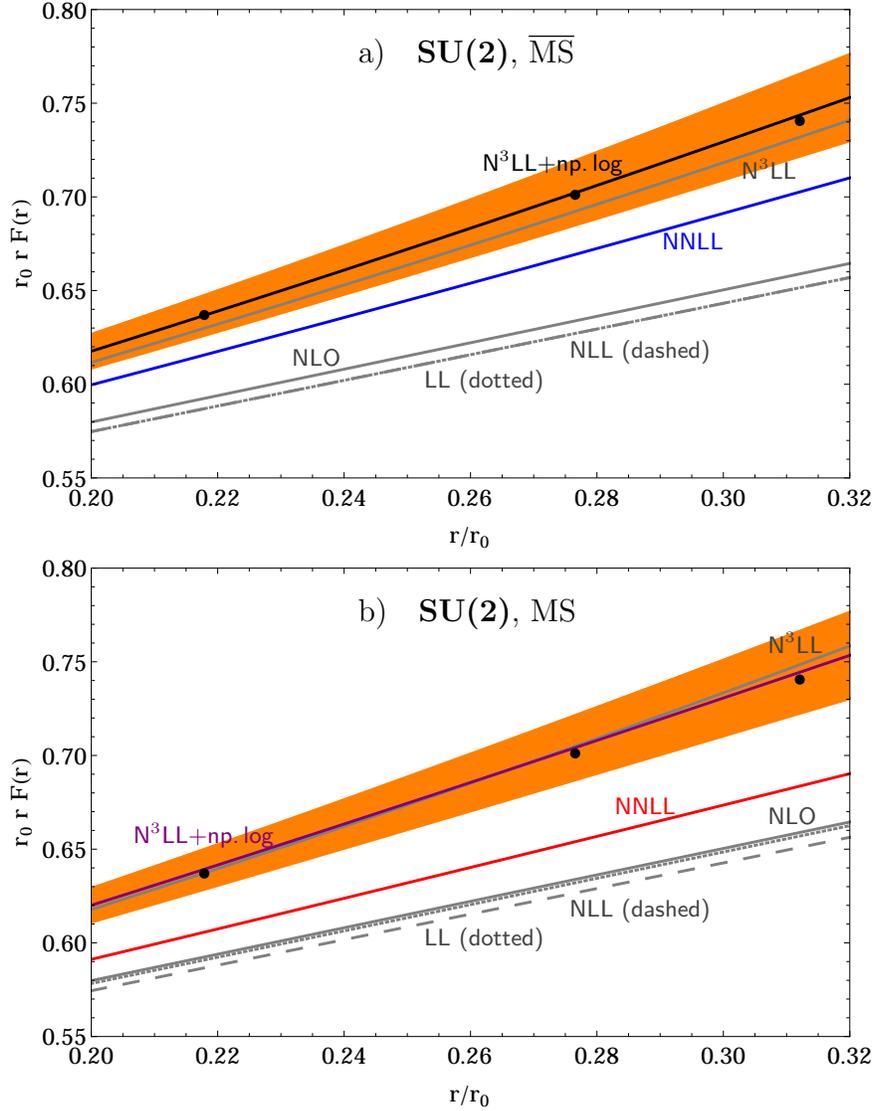

The value of $\al$ to be used in our perturbative results for a comparison with the SU(3) data is given by $\al = \frac{0.18}{r_0}$~\cite{Luscher:2002qv}, where $r_0=0.5\,fm$ is the Sommer scale. 
The corresponding SU(2) value for $\al$ was obtained in Ref.~\cite{HariDass:2007tx} from a fit using a wrong ansatz to parametrize the corrections higher than LO. We repeated the fit, as described in Ref.~\cite{HariDass:2007tx}, using the analytic result according to Eq.~(\ref{NNLL}) in the scheme for which $d_X=1$ (i.e. $c_X=e^{\frac12(\gamma_E + \ln \pi)}$) and find $\al=\frac{0.29}{r_0}$.
This number is quite close to the one of Ref.~\cite{HariDass:2007tx} ($\frac{0.30}{r_0}$) and indicates that the multipole expansion works quite well for the available SU(2) short distance lattice data. 

The numerical values $\al r_0=0.18$ for SU(3) and $\al r_0=0.29$ for SU(2) suggest that for $r<r_0$ the multipole expansion should work well. 
This is confirmed in Fig.~\ref{Plots}a) for SU(2) and in Fig.~\ref{Plots}b) for SU(3), where we plot the force $F(r)$ (normalized to a dimensionless number by factors of $r_0$) as a function of $r/r_0$. 
We also display the lattice points of Refs.~\cite{HariDass:2007tx} and~\cite{Luscher:2002qv} (black dots) for comparison.
In these figures we observe a convergent sequence of the LO, NLO and NNLO. 
At the latter order several curves are shown in Fig.~\ref{Plots}a/b), as the result depends on the order at which the ultrasoft expansion is truncated. Yet, the convergent pattern of the multipole expansion remains irrespectively of which NNLO result is used.
Another way to note this is to look at the difference between the NLO curve and the lattice points, which should account for all corrections beyond NLO and is smaller than the difference between the LO and NLO result in the plotted range.

We now turn to the study of the ultrasoft expansion. The $\al/\Delta V$ expansion does not work that well, 
even for the shortest distances that were probed on the lattice. At
$r/r_0 \simeq 0.22$ for SU(2) and $r/r_0 \simeq 0.35$ for SU(3) we have 
$\big[C_A \al / \Delta V_{\msb} \big]_{SU(2)} \simeq 0.60$ and 
$\big[C_A \al / \Delta V_{\msb} \big]_{SU(3)} \simeq 0.69$, respectively. 
In the $\rm MS$ scheme it is even worse, since generally $\Delta V_{\rm MS}<\Delta V_{\msb}$. The LL and NLL contributions are 
very small (maybe anomalously small), much smaller than the NNLL contribution, which, however, improves the agreement 
with the lattice data points. The pattern is somewhat similar in the $\rm MS$ and $\msb$ 
scheme despite the fact that $C_A \al/\Delta V_{\msb}<C_A \al/\Delta V_{\rm MS}$.
We illustrate this for the SU(2) case in Figures~\ref{Plotsshort}a) and b), where we show the (normalized) results for $r F(r)$ in a short distance range in the $\MS$ and MS scheme, respectively. The $F(r)$ plots in Fig.~\ref{Plots}a) and Fig.~\ref{Plots}b) also display NNLL MS and $\MS$ curves up to $r=1.5\, r_0$ for SU(2) and SU(3), respectively.
In addition to the plotted scheme dependent results of Eq.~(\ref{Afinal}) we have also studied the convergence of Eq.~(\ref{Bfinal}). 
We find a qualitatively similar pattern as for the MS and $\msb$ scheme through NNLL.

At N$^3$LL the constant $c_{2,0}$ appears in the analytic expressions and at N$^4$LL $c_{3,0}$ in addition.
They are both unknown at present. Since the convergence in the 
$\al/\Delta V$ expansion is not good enough, we do not attempt to extract $c_{3,0}$ from the comparison with the available lattice data and focus on $c_{2,0}$.
The strategy we follow is to perform one-dimensional fits of our analytic expressions to the data at the shortest available distances. There is some scheme dependence in this determination, which we exploit to estimate the perturbative error of the fitted parameter $c_{2,0}$.
On the one hand we use Eq.~(\ref{Afinal}) {\bf (A)} and on the other hand Eq.~(\ref{Bfinal}) {\bf (B)} as the perturbative input for the fits, both in the $X=\rm MS$ as well as the $X=\MS$ scheme. 
In all cases we fit with N$^3$LL as well as with N$^4$LL precision, excluding however all non-logarithmic terms at N$^4$LL order, since these include the unknown $c_{3,0}$. We refer to the latter approximation as N$^3$LL+np.\,log.
We also vary the number of lattice data points in the fit to check the reliability of the results.
The numerical value should not depend much on the number of data points if we are at short enough distances. 
Last but not least, we carry out separate fits for $N_c=2$ and $N_c=3$, as we expect $c_{2,0}$ to be roughly independent of $N_c$ (for $n_f=0$). 
The difference in the fit values for $c_{2,0}$ then allows to draw conclusions on the fit quality.

\begin{table}[ht]
\begin{tabular}{|l|c|c|c|c|}
\hline
gauge group: & SU(2) & SU(2) & SU(2) & SU(3)\\
\hline
\# of fit points: & 1 & 3 & 5 & 1 \\ 
\hline
\hline
N$^3$LL+np.\,log {\bf (A)}: $\msb$ fit & $-1.96$ & $-2.10$ & $-2.19$ & $-1.17$ \\
\hline
N$^3$LL+np.\,log {\bf (A)}: $\msb$ from $\rm MS$ fit & $-2.31$ & $-2.23$ & $-2.02$ & $-1.20$ \\
\hline
N$^3$LL+np.\,log {\bf (B)}: $\msb$ fit & $-2.93$ & $-3.24$ & $-3.46$ & $-2.37$ \\
\hline
N$^3$LL {\bf (A)}: $\msb$ fit & $-1.00$ & $-1.26$ & $-1.50$ & $-0.38 $ \\
\hline
N$^3$LL {\bf (A)}: $\msb$ from $\rm MS$ fit & $-2.03$ & $-2.37 $ & $-2.68 $ & $-1.52$ \\
\hline
N$^3$LL {\bf (B)}: $\msb$ fit & $-0.70$ & $-0.99$ & $-1.30$ & $-0.13$ \\
\hline
\hline
N$^3$LL+np.\,log {\bf (A)}: $\rm MS$ fit & $+0.37$ & $+0.45$ & $+0.66$ & $+1.48$ \\
\hline
N$^3$LL+np.\,log {\bf (A)}: $\rm MS$ from $\msb$ fit & $+0.73$ & $+0.58$ & $+0.49$ & $+1.51$ \\
\hline
N$^3$LL+np.\,log {\bf (B)}: $\rm MS$ fit & $-0.25$ & $-0.55$ & $-0.78$ & $+0.32$ \\
\hline
N$^3$LL {\bf (A)}: $\rm MS$ fit & $+0.65$ & $+0.31$ & $-0.00$ & $+1.16$ \\
\hline
N$^3$LL {\bf (A)}: $\rm MS$ from $\msb$ fit & $+1.69 $ & $+1.42$ & $+1.18$ & $+2.30$ \\
\hline
N$^3$LL {\bf (B)}: $\rm MS$ fit & $+1.99$ & $+1.69$ & $+1.38$ & $+2.55$ \\
\hline
\end{tabular}
\caption{\it Results for $c_{2,0}$ from fitting the force derived from 
{\bf (A)}: Eq. (\ref{Afinal}) truncated at N$^3$LL or N$^3$LL+np.\,log,  
and {\bf (B)}: Eq. (\ref{Bfinal}) truncated at N$^3$LL or N$^3$LL+np.\,log, to lattice 
data for $N_c=2$ and $N_c=3$ in the $\rm MS$ and $\MS$ scheme. 
The (first/first three/first five) lattice points with smallest distances $r$ have been used for the fit. 
To obtain the values in the rows labeled ``\,$\msb$ from $\rm MS$ fit'' and ``\,$\rm MS$ from $\msb$ fit'' we have transformed the {\rm MS} fit values to the $\MS$ scheme and vice versa, by adding/subtracting the difference coming from the scheme ($X$) dependent terms in Eq.~(\ref{FinalNNNLL}).
Note that for option {\bf (B)} such a scheme transformation reproduces the same results as from the direct fits. Therefore, there is no uncertainty from the scheme dependence in this case.
\label{c20}} 
\end{table}

Table~\ref{c20} lists the numbers for $c_{2,0}$ resulting from the fits to the data points at 
the smallest available distances $r$ in the renormalization schemes $\rm MS$ and $\msb$, using method {\bf (A)} and  {\bf (B)}, and truncating the perturbative expression at N$^3$LL and at N$^3$LL+np.\,log, respectively. 
We should not equally weight all these determinations. First, we expect method ${\bf (A)}$ to yield better results than method ${\bf (B)}$, since an extra resummation has been performed. Within method ${\bf (A)}$ we believe that the $\msb$ fit is more reliable, since $\Delta V_{\msb}>\Delta V_{\rm MS}$. 
We also expect that increasing the number of included data points (with larger $r$) decreases the fit quality. 
Therefore, we take the $\MS$ ${\bf (A)}$ fit (N$^3$LL+np.\,log) to the first lattice point at $r/r_0 \simeq 0.22$, where we expect the smallest theoretical error from the truncation of the perturbation series, as our central value 
($c_{2,0}^{\rm MS}=c_{2,0}^\msb+2.68$): 
\be
\label{c20fit}
c_{2,0}^\msb =  -1.96 \pm 1.5\,, \qquad c_{2,0}^{\rm MS} = 0.73 \pm 1.5.
\ee
We conservatively estimate the (symmetric) error such that all the SU(2) N$^3$LL+np.\,log and N$^3$LL determinations in Table~\ref{c20} are included in the error band.
Let us now examine the anatomy of the uncertainty in more detail: 
We first notice that the fitted value for $c_{2,0}$ does not depend much on the number of data points, with variations $\sim 0.5$ at most. 
Truncating at N$^3$LL or at N$^3$LL+np.\,log produces variations $\sim 1$ at most, whereas using method {\bf (A)} or {\bf (B)} produces variations of similar size. Finally, we find it quite reassuring and nontrivial that changing from the MS to the $\msb$ scheme (and viceversa) generates perfectly compatible numbers. The same is true if we compare SU(2) and SU(3) determinations. 

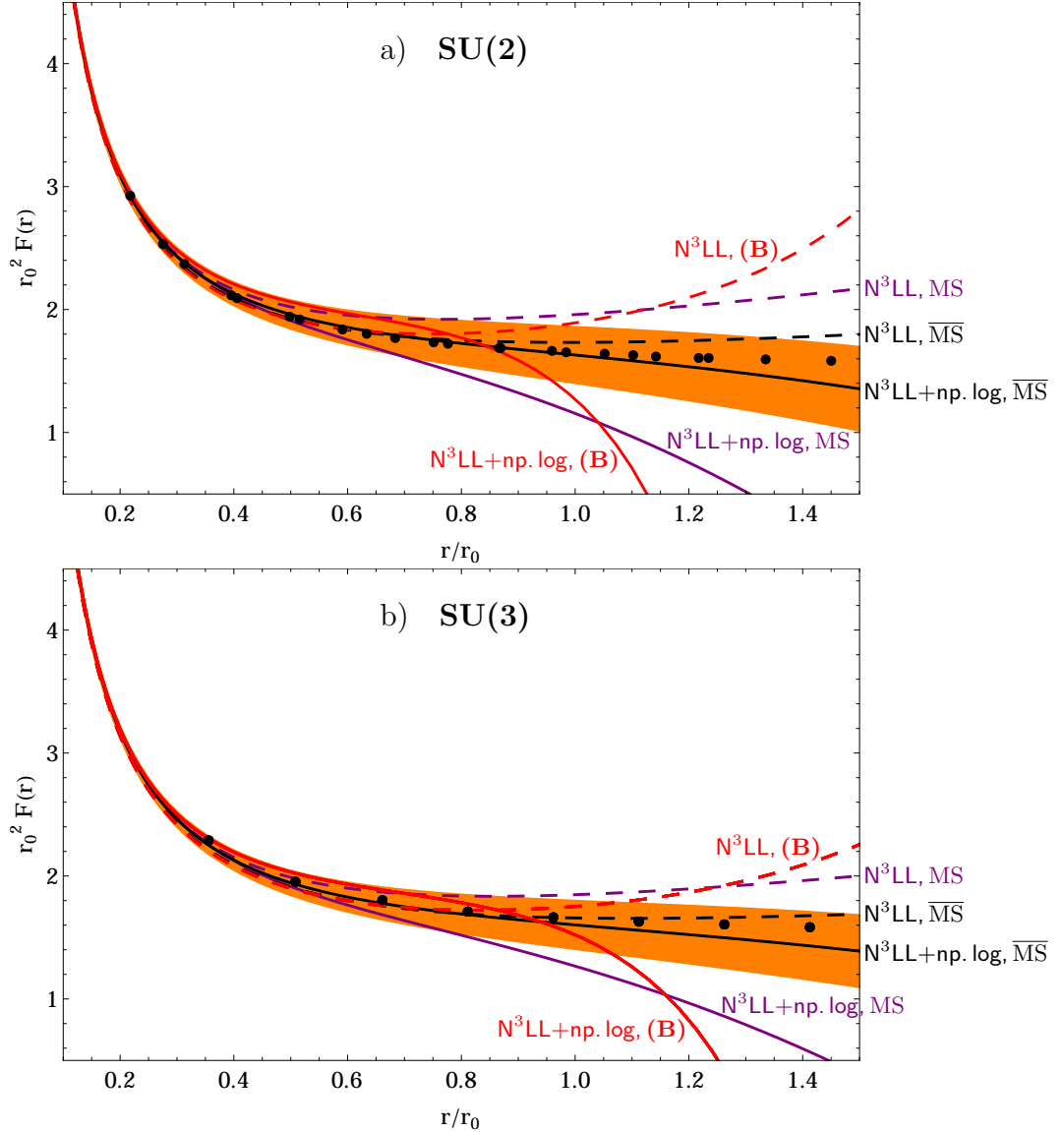
\begin{figure}[ht]
\includegraphics[width=0.7 \textwidth]{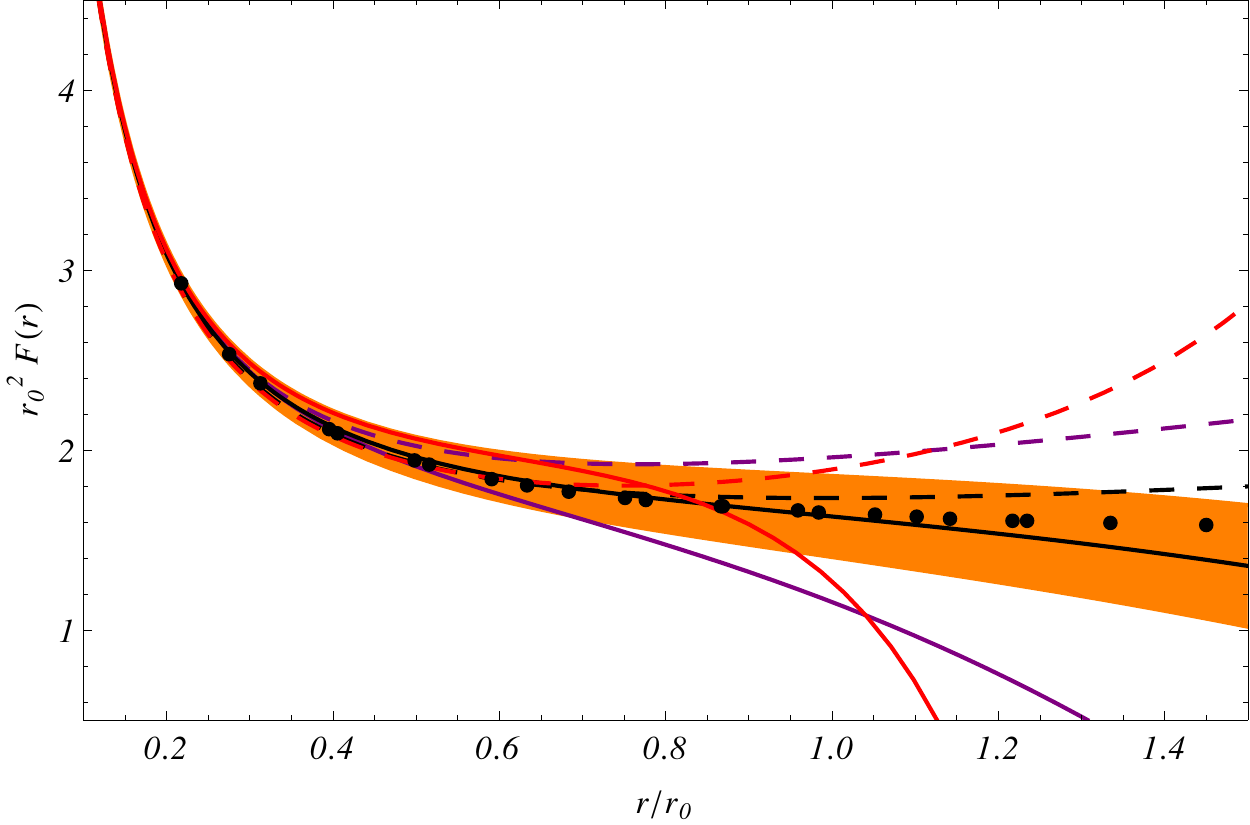}
\put(-185,195){\large a)\quad \bf SU(2)}
\put(-75,45){\sf \color{violet} N$^3$LL$+$np.\,log,\,\rm MS}
\put(1,104){\sf \color{violet} N$^3$LL,\,\rm MS}
\put(1,86){\sf N$^3$LL,\,$\msb$}
\put(1,64){\sf N$^3$LL$+$np.\,log,\,\rm $\msb$}
\put(-166,38){\sf \color{red} N$^3$LL$+$np.\,log,\,\bf(B)}
\put(-72,119){\sf \color{red} N$^3$LL,\,\bf(B)}

\includegraphics[width=0.7 \textwidth]{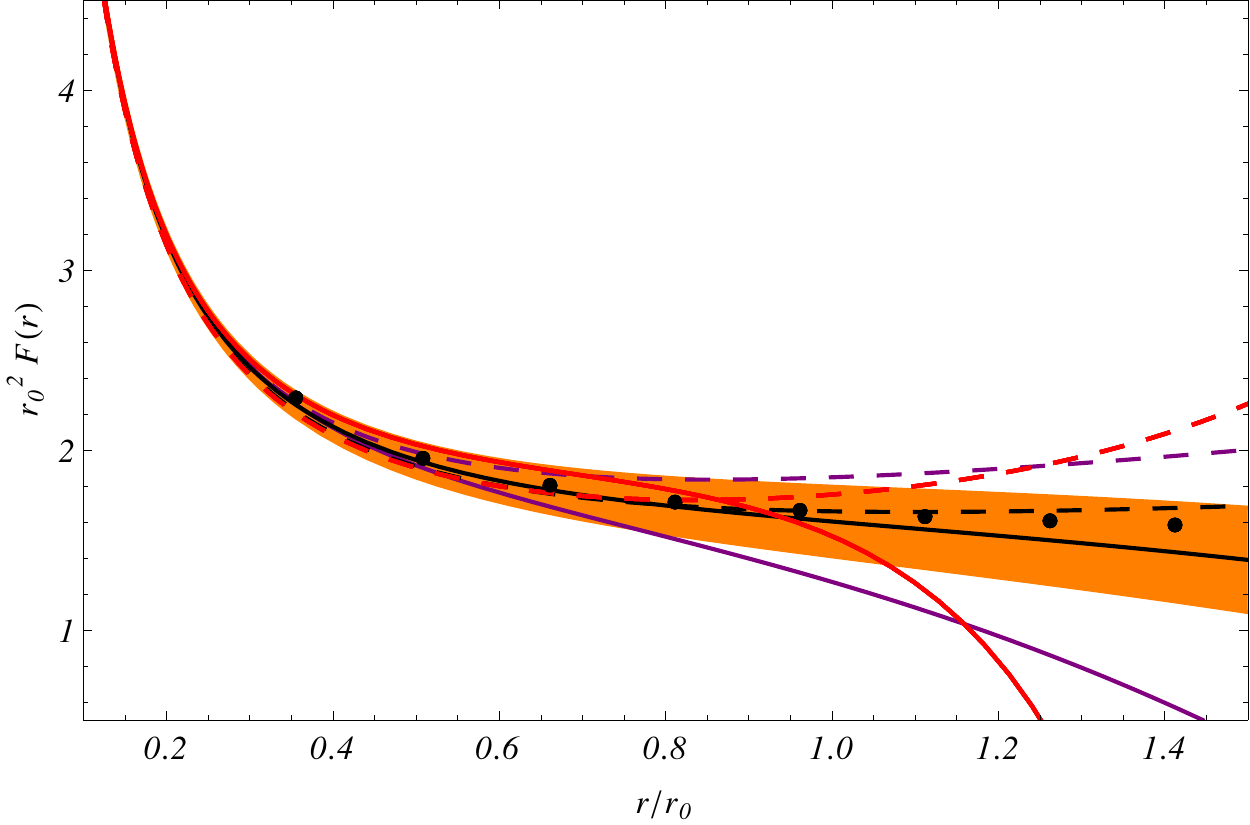}
\put(-185,195){\large b)\quad \bf SU(3)}
\put(-54,46){\sf \color{violet} N$^3$LL$+$np.\,log,\,\rm MS}
\put(1,96){\sf \color{violet} N$^3$LL,\,\rm MS}
\put(1,81){\sf N$^3$LL,\,$\msb$}
\put(1,66){\sf N$^3$LL$+$np.\,log,\,\rm $\msb$}
\put(-140,37){\sf \color{red} N$^3$LL$+$np.\,log,\,\bf(B)}
\put(-56,107){\sf \color{red} N$^3$LL,\,\bf(B)}

\caption{\it Plots of the NNLO analytic results for the force $F(r)$ in $r_0$ units with $n_f=0$ in comparison to the available lattice data (black dots):
a) $N_c=2$. We display the N$^3$LL and the N$^3$LL$+$np.\,log curves in the MS and $\msb$ scheme according to Eq.~(\ref{Afinal}) {\bf (A)} and according to Eq.~(\ref{Bfinal}) {\bf (B)}. The (orange) band visualizes the error of the N$^3$LL$+$np.\,log $\msb$ expression due to the uncertainty of $c_{2,0}$ in Eq.~(\ref{c20fit}).
b) Same curves as in panel~a), but with $N_c=3$.
\label{PlotsUS}
}
\end{figure}

We adopt the $c_{2,0}$ values in Eq.~(\ref{c20fit}) for the N$^3$LL and N$^3$LL+np.\,log curves shown in Figs.~\ref{Plots}, \ref{Plotsshort}, and \ref{PlotsUS}.
In other words we have fitted the N$^3$LL+np.\,log curve in the $\msb$ scheme to the first data point for $N_c=2$.
Hence this point lies exactly on the $\msb$ curve, whereas in Fig.~\ref{Plots}b), where $N_c=3$ and the same value for $c_{2,0}^\msb$ 
has been used, it does not. 
For all $\msb$ N$^3$LL+np.\,log curves in Figs.~\ref{Plots}, \ref{Plotsshort}, and \ref{PlotsUS} and for the MS N$^3$LL+np.\,log curve in Fig.~\ref{Plotsshort}~b) we also display error bands (orange) according to the uncertainty of $c_{2,0}$ as given in Eq.~(\ref{c20fit}).

The good agreement between prediction and data beyond the first data point is nontrivial.
This is especially the case for the (solid black) $\MS$ result, which almost perfectly agrees with the lattice data even slightly beyond $r_0$, as we can see in Fig.~\ref{PlotsUS}.
The figure also reveals that the difference between the $\msb$ N$^3$LL and N$^3$LL+np.\,log curves is relatively small up to $r \sim r_0$, which we find quite reassuring. 
The MS results and the expressions from method ${\bf (B)}$ break down before (around $0.7\,r_0$ and $r_0$, respectively) as expected. 
Nevertheless, even these results describe the lattice data reasonably well over a large range of~$r$.

In Fig.~\ref{Plotsshort} we focus on the shortest available lattice data and plot all known terms of the ultrasoft expansion. If we take into account the LL and NLL lines the convergence is not good. Nevertheless, in the $\MS$ scheme, we find a convergent sequence if we look at the NLL, NNLL, N$^3$LL and 
N$^3$LL+np.\,log curves. The convergence pattern in the MS scheme is worse though the final result also agrees well with the lattice points.
Working with Eq.~(\ref{Bfinal}) produces curves that converge slightly worse than in the $\MS$ scheme, while the N$^3$LL+np.\,log result is a bit off the data points.

From the above analysis, we conclude that the lattice simulations so far have produced data at short enough 
distances to quantitatively test the multipole ($\al r$) expansion, cf. Figs.~\ref{Plots}a) and~b). Results for the $\al/\Delta V$ expansion are less 
conclusive but encouraging, in particular in the $\MS$ scheme.

\section{pNRQCD(D=2)}
\label{secD2}

It is also interesting to investigate the static quark-antiquark potential in two space-time dimensions.
In the following we set $n_f =0$.
For $D=2$ the exact result for the singlet static potential is known. It can be derived from its standard (non-perturbative) definition via a Wilson-loop:
\be
E_s(r) := \lim_{T \to \infty} \frac{i}{T} \ln\, \langle {\frac1{N_c} \rm Tr} \,{\cal P} \exp\big(-i g \oint_{\Gamma} A_\mu dx^\mu\big) \rangle\,,
\label{Wilsonloop}
\ee
where the contour $\Gamma$ of the integral in Eq.~(\ref{Wilsonloop}) is a rectangle with spatial extension $r$ and extension $T$ along the time coordinate axis in Minkowski space. Choosing axial ($A_1 \equiv 0$) gauge in exactly two dimensions, the only non-vanishing components of the gluon field-strength tensor are $F_{01}=-F_{10}=-\partial_1 A_0$. Hence, ($A_0$) gluons do neither interact among themselves, nor propagate in time, since no time derivative is left in the Lagrangian. 
Therefore only planar ``ladder'' diagrams (with ``potential'' gluons) contribute to the Wilson-loop. 
In such a quasi-Abelian situation, one can forget about the path-ordering operator 
${\cal P}$ in Eq.~(\ref{Wilsonloop}) and treat the SU($N_c$) generators $T^a$ in the exponential 
as if they were proportional to the unit matrix. Finally we arrive at the result for the 
corresponding Abelian Wilson-loop multiplied by $\frac1{N_c}{\rm Tr}\, T^aT^a=C_F$. This result 
is identical to the perturbative tree-level contribution
\be
E_s(r) = 2 \pi C_F \al\, r
\label{ExactPot}
\ee
and therefore proving exponentiation.

The computation in pNRQCD is organized differently, because of the factorization of the different scales that contribute 
to the potential. For exactly $D=2$ pNRQCD recovers Eq. (\ref{ExactPot}), since the soft contribution is precisely Eq. 
(\ref{ExactPot}) and there are neither ultrasoft nor non-perturbative gluons. For $D > 2$ one would expect the result to 
converge to Eq. (\ref{ExactPot}), when $D \rightarrow 2$, provided the limit is smooth. 
The soft two-loop calculation \cite{Schroder:1999sg} yields however
\be
V_s(r) = 2 \pi C_F  \al\, r \,-\, 2  \pi C_A C_F \al^2 r^3  \,+\, \frac{\pi}{27} (33+2 \pi ^2) C_A^2 C_F \al^3 r^5
\label{Vsoft2D}
\ee
for $D \to 2$ and $n_f =0$. 
From the effective theory viewpoint, we expect that the ultrasoft contributions to the static energy, 
calculated in pNRQCD for $D \to 2$, cancel the $\al^2$ and $\al^3$ terms in Eq.~(\ref{Vsoft2D}). Though, checking this is a delicate issue, 
since as mentioned in subsection~\ref{RenAndGen}
the proper ultrasoft expansion parameter is now $\frac{\Delta V}{g} \sim g r$, i.e. non-perturbative. Thus the intrinsically 
non-perturbative ultrasoft ($\Delta V/g$) expansion and the perturbative multipole ($g r$) expansion mix.

The non-perturbative nature of the ultrasoft expansion does not allow a calculation in terms of pNRQCD loop diagrams for $D \not= 2$. 
Expressions like Eq.~(\ref{deltaVUS}) should rather be expanded for small $\Delta V$ and the full non-perturbative result 
for the chromoelectric correlator should be used. Of course the latter is not known in dimensional 
regularization. 

Despite that we have observed interesting cancellations. The ultrasoft one-loop contribution at the 
lowest non-vanishing order in the multipole expansion, Eq.~(\ref{USbare1loop}), gives
\be
\delta E^{us}({\rm 1-loop}) = 2 \pi C_A C_F  \al^2 r^3 
\ee
in the limit $D\to2$ and exactly cancels the soft $\al^2$ contribution in Eq.~(\ref{Vsoft2D}).

At this point we are unfortunately not able to check the (anticipated) cancellations at higher orders in $\al$, as 
they may require higher order terms in the multipole expansion. We are also not sure about the relevance of the above result, as 
it follows from a perturbative ultrasoft computation. 
Nevertheless the observed cancellation could be interpreted as another check of the applied effective field theory 
approach, which uses perturbative methods to derive meaningful results for the singlet static energy in arbitrary dimensions.

\section{Conclusions}
\label{conclusion}

The main achievement of this paper is a precise determination of the static energy 
of a color singlet quark-antiquark system in three space-time dimensions for $\al\, r \ll 1 $. 
We have complemented the strictly perturbative (soft) two-loop QCD result~\cite{Schroder:1999sg} 
by an effective theory calculation in pNRQCD to account for effects at the ultrasoft scale 
($\Delta V=V_o\!-\!V_s$), which are not accessible by pQCD.

We observe a perfect cancellation of the IR singularities in the soft pQCD result at $\ord(\al^3r^2)$ 
and the UV divergences in the one- and two-loop pNRQCD corrections, 
i.e. at NLL and NNLL order in the ultrasoft $\al/\Delta V$ expansion. This represents an 
independent and nontrivial check of the soft pQCD and the ultrasoft pNRQCD contributions, 
because the singlet static energy is a physical observable and must be IR finite. 
We also observe a cancellation of logarithmic divergences that can be associated with the 
leading renormalons of the potential and the quark mass in four dimensions.

Demanding complete IR finiteness we have determined indirectly the UV divergences at ultrasoft 
three-loop level, without having calculated the actual diagrams. In fact the only missing piece in 
our N$^3$LL prediction for the singlet static energy up to $\ord(r^2)$, Eq.~(\ref{FinalNNNLL}), 
is the ultrasoft three-loop constant $c_{2,0}$. We also provide expressions for the N$^4$LL correction 
and the known leading non-perturbative contribution from the gluon 
condensate~\cite{Schroder:2003uw,DiRenzo:2006nh,Hietanen:2004ew,Hietanen:2006rc}, 
which is parametrically of the same order. They are complete up to the ultrasoft four-loop constant 
$c_{3,0}$. Thus the precision is currently limited by missing perturbative three- and four-loop computations.

We have compared our theoretical predictions with lattice data. We have found that 
the $\al r$ (multipole) expansion is already tested by present lattice simulations. For the $\al/\Delta V$ expansion the situation is less conclusive but promising, especially in the $\MS$ scheme: 
We are able to extract a value for $c_{2,0}$, albeit with large uncertainties, and agreement with lattice data is obtained up to $r \sim r_0$. 

With lattice data at smaller distances and analytic results for $c_{2,0}$ and $c_{3,0}$, 
an interesting cross-check of the three-dimensional lattice simulation for the gluon 
condensate~\cite{Hietanen:2004ew,Hietanen:2006rc} in Eq.~(\ref{eq:condensate}) would be feasible.

We have also studied the two dimensional case, for which the exact result is 
known and equivalent to the tree-level result. Therefore, strong cancellations have to 
occur among the contributions from the different scales: soft, ultrasoft and $g$ in the $D \rightarrow 2$ limit.
We have found a cancellation between the one-loop soft and the one-loop ultrasoft results in dimensional regularization for $D \rightarrow 2$.
The relevance of this cancellation is however not clear to us, because it follows from a perturbative ultrasoft computation. 
Nevertheless, we think that it could be worthwhile to explore the $D \rightarrow 2$ limit 
in more detail, and see whether it could provide nontrivial information on the structure of the 
perturbation series for general dimensions. We leave this issue open in this paper.

\acknowledgments{
This work was partially supported by the EU network contract
MRTN-CT-2006-035482 (FLAVIAnet), by the Spanish 
grant FPA2007-60275 and by the Catalan grant SGR2009-00894.
}

\end{document}